\documentclass[fleqn,usenatbib]{mnras}

\usepackage{amsmath,amssymb,txfonts,makecell}
\usepackage{hyperref}
\usepackage{cleveref}
\usepackage{graphicx}
\usepackage{tabularx,diagbox}
\usepackage{soul,color}
\usepackage{ulem}

\renewcommand{\thesection}{\arabic{section}}

\newcommand{\diff}{\mathrm{d}}

\crefname{equation}{eq.}{eqs.}
\crefname{figure}{Fig.}{Fiqs.}

\def\be{\begin{equation}}
\def\ee{\end{equation}}
\def\ba{\begin{align}}
\def\ea{\end{align}}

\def\kmsMpc{\ensuremath\,\rm{km}\,\rm{s}^{-1}/\rm{Mpc}}

\setstcolor{red}
\setul{-.5ex}{.3ex}

\title[]{{Exploring the evidence for a large local void with supernovae Ia data}}

\author[V. V. Lukovi\'{c} et al.]{
    Vladimir V. Lukovi\'{c},$^{1,2}$\thanks{E-mail: vladimir.lukovic@roma2.infn.it}
    Balakrishna S. Haridasu,$^{1,2}$\thanks{E-mail: haridasu@roma2.infn.it}
    and Nicola Vittorio$^{1,2}$\thanks{E-mail: nicola.vittorio@roma2.infn.it}
    \\
    $^{1}$Dipartimento di Fisica, Universit\`{a} di Roma "Tor Vergata", Via della Ricerca Scientifica 1, I-00133, Roma, Italy\\
    $^{2}$Sezione INFN, Universit\`{a} di Roma "Tor Vergata", Via della Ricerca Scientifica 1, I-00133, Roma, Italy
}

\date{Accepted 2019 October 25. Received 2019 October 22; in original form 2019 July 25}

\pubyear{2019}

\begin{document}
    \label{firstpage}
    \pagerange{\pageref{firstpage}--\pageref{lastpage}}
    \maketitle
    \begin{abstract}
In this work we utilise the most recent publicly available type Ia supernova (SN Ia) compilations and implement a well formulated cosmological model based on Lema\^{i}tre-Tolman-Bondi metric in presence of cosmological constant $\Lambda$ ($\Lambda$LTB) to test for signatures of large local inhomogeneities at $z\leq0.15$. Local underdensities in this redshift range have been previously found based on luminosity density (LD) data and galaxy number counts. Our main constraints on the possible local void using the Pantheon SN Ia dataset are: redshift size of $z_{\rm size}=0.068^{+0.021}_{-0.030}$; density contrast of $\delta\Omega_0/\Omega_0=-10.5_{-7.4}^{+9.3}\%$ between 16th and 84th percentiles. Investigating the possibility to alleviate the $\sim9\%$ disagreement between measurements of present expansion rate $H_0$ coming from calibrated local SN Ia and high-$z$ cosmic microwave background data, we find large local void to be a very unlikely explanation alone, consistently with previous studies. However, the level of matter inhomogeneity at a scale of $\sim$100Mpc that is allowed by SN Ia data, although not expected from cosmic variance calculations in standard model of cosmology, could be the origin of additonal systematic error in distance ladder measurements based on SN Ia. Fitting low-redshift Pantheon data with a cut $0.023<z<0.15$ to the $\Lambda$LTB model and to the Taylor expanded luminosity distance formula we estimate that this systematic error amounts to $1.1\%$ towards the lower $H_0$ value. A test for local anisotropy in Pantheon SN Ia data yields null evidence. Analysis of LD data provides a constraint on contrast of large isotropic void $\delta\Omega_0/\Omega_0=-51.9\%\pm6.3\%$, which is in $\sim4\sigma$ tension with SN Ia results. More data are necessary to better constrain the local matter density profile and understand the disagreement between SN and LD samples.
    \end{abstract}
    
    
    \section{Introduction}
    \label{sec:intro}
    The most reliable standard candles and one of the fundamental low-redshift astrophysical observables \citep{Betoule14,Scolnic17}, type Ia supernovae (SN Ia), also provide a cosmology-independent measurement of present expansion rate through the local distance ladder \citep{Riess16} (hereafter R16).
    In concert with observations of baryon acoustic oscillation (BAO) peak from galaxy redshift surveys \citep{Eisenstein05,Aubourg15,Alam17,Bautista17,MasdesBourboux17,Zhao19} and cosmic microwave background (CMB) radiation data \citep{WMAP13,Planck18L}, SN Ia contributed to establishment of the standard model of cosmology.
    Astonishing concordance between different cosmological observables brought by technical advancements in the last decades was followed by a number of inconsistent and/or unexpected results, e.g. 
    estimated values of matter clustering parameter, $S_8$, from low and high redshifts \citep{Planck16SZ,Joudaki19,Lange19,Martinelli19}, CMB parameter constraints in curvature extension of base $\Lambda$ cold dark matter ($\Lambda$CDM) model \citep{Planck18} (hereafter P18), as well as several unexplained CMB anomalies \citep{Schwarz16}, etc.
    The best-known amongst $\Lambda$CDM tensions and one of the biggest challenges in modern cosmology is related to the $H_0$ value.
    Defining the scale of extra-galactic distance through the Hubble radius, as well as the present expansion rate, $H_0$ is a fundamental cosmological parameter both for CMB and low-redshift data \citep{Freedman01,Freedman10}.
    The most stringent $\Lambda$CDM estimate coming from P18, $H_0=67.36\pm0.54\kmsMpc$, is in $4.4\sigma$ tension with the most recent local measurement based on calibrated SN Ia, $H_0= 74.03 \pm 1.42\kmsMpc$, by \cite{Riess19} (hereafter R19).
    In comparison, $H_0$ measurements from the time delay of strongly lensed quasars support a higher value \citep{Birrer19,Wong19}; using differential age method with cosmic chronometers \citep{Jimenez02} provides an estimate in the middle of two results in tension \citep{Jimenez19,ChenRatra17,Lukovic16}; while the constraints coming from gravitational waves are still fairly weak \citep{Liao17}.
    Although the quality of BAO data has substantially improved over the last decade, providing some of the most stringent cosmological constraints on expansion dynamics \citep{Haridasu17a, Ramanah19, Zhang19}, their $H_0$ estimate is degenerated with the sound horizon parameter, whose value relies on CMB observations and early Universe physics \citep{Macaulay19, Lemos19, Aubourg15, Cuceu19}.
    
    The intercept of SN Ia magnitude-redshift relation is determined by their absolute magnitude ($M_B$) and the Hubble radius, resulting in two degenerate parameters in SN Ia analyses.
    While the high-redshift SN Ia are necessary for assessing the underlying physics of dark energy, the low-redshift end ($z\leq0.15$) of SN Hubble diagram can be analysed with minimal assumptions about properties of the cosmic fluid and expansion dynamics related only to the well-accepted cosmic acceleration \citep{Haridasu17}.
    Indeed, calibrating SN Ia absolute magnitude with the use of Cepheid variable stars and fixing the local distance-redshift relation (at $z\leq0.15$) leads to a remarkable cosmology-independent measurement of $H_0$
    \citep{Riess05, Riess07, Riess09, Riess11, Riess16, Riess18}.
    Different calibration techniques consider the usage of strong gravitational lensing \citep{Taubenberger19}, tip of the red giant branch \citep{Beaton16,Freedman19}, H{\footnotesize II} regions \citep{Arenas18}, etc.
    Due to the strong correlation of cosmic expansion dynamics and $H_0$ value, the model-independent local estimates remain crucial for the study of late-time cosmic evolution [see also \cite{Tutusaus19, Haridasu18_GP}].
    
    As is well-known, a large inhomogeneity in local matter density distribution can affect the geometry of local Universe and the distance measures, consequently biasing the resulting model inferences and the $H_0$ estimate. 
    A few independent groups used galaxy survey catalogues to test the local matter distribution, reporting hints of a large local void \citep{Keenan12, Keenan13, Whitbourn14, Whitbourn16,Boehringer19}.
    Concretely, \cite{Keenan13} (hereafter KBC13) used UKIDSS Large Area Survey \citep{Lawrence07}, Galaxy and Mass Assembly (GAMA) \citep{Driver10,Driver11} and 2M++ catalogue \citep{Lavaux11} to construct luminosity density sample (hereafter LD) over the redshift range $0.01<z<0.2$.
    Relating it to the total matter density in the local Universe, they find evidence for an underdense region and suggest a void of about $300{h_{70}}^{-1}$Mpc in size and $-30\%$ contrast with respect to the background.
    \cite{Whitbourn14} (hereafter WS14) used 6dFGS, SDSS, and GAMA surveys to examine the local galaxy density field over 20\% of the sky and up to $450{h_{70}}^{-1}$Mpc in depth.
    Both number counts and peculiar velocity fields show evidence for at least $200{h_{70}}^{-1}$Mpc large local underdensity with $-15\%\pm3\%$ average contrast, ranging between $-40\%$ and $-5\%$ in different directions.
    Analysing a catalogue of 1653 galaxy clusters detected by their X-ray emission \cite{Boehringer19} report evidence for $\sim140$ Mpc large local underdensity with density contrast of $20\%\pm10\%$.
    
    The results of KBC13, WS14 and \cite{Boehringer19} go against the standard cosmological paradigm and the assumption of large-scale homogeneity.
    In fact, indications for a few hundred Mpc large local underdensity encouraged renewed popularity of {\it void models} based on Lema\^{i}tre-Tolman-Bondi (LTB) metric, also motivated as a remedy for the $H_0$ tension \citep{Moffat16,Tokutake17,Hoscheit17,Shanks19,Kenworthy19} and even as a possible explanation for the CMB cold spot \citep{Szapudi15}.
    Contrasting the homogeneous and isotropic Friedmann-Lema\^{i}tre-Robertson-Walker (FLRW) models, the LTB metric \citep{Lemaitre27,Lemaitre33,Tolman34, Bondi47} describes an isotropic but inhomogeneous dust system, which on cosmological scales can be used for studying cold dark matter density distribution.
    Since the discovery of cosmic acceleration, LTB metric was used to construct toy models that challenge the cosmological constant paradigm with an alternative scenario in which the apparent accelerated expansion is a result of the strongly underdense local Universe that smoothly converges to Einstein de Sitter solution on higher redshifts  \citep{Celerier00,Alnes06a,Clifton08,February10,Nadathur11,Bolejko11,Zhang15,Stahl16}.
    Since the deceleration-acceleration transition redshift is well constrained at $z_t\approx0.6$ [see e.g. \cite{Haridasu18_GP,Gomez-Valent18, Mukherjee19} for model-independent estimates], this alternative explanation requires a giant ($\,\approx\,$3Gpc) isotropic void that proved to be very unlikely compared to the standard cosmological model \citep{Zibin11,Vargas17,Amendola13,Lukovic16}.
    However, extending the LTB model with the addition of cosmological constant $\Lambda$, one can describe large (order of $100$Mpc) local inhomogeneous matter distribution inside a $\Lambda$CDM background \citep{Valkenburg12,Valkenburg14,Rigopoulos12}.
    
    While the level of cosmic variance in matter distribution expected for standard model can only partially relieve the Hubble constant problem \citep{Marra13,Wu17,Camarena18}, \cite{Tokutake17} found that a strongly underdense local inhomogeneity with non-vanishing cosmological constant, modeled with an approximate numerical solution for LTB metric, can fully coincide the CMB constraints with the local $H_0$ measurement.
    Similarly, \cite{Hoscheit17} (hereafter HB17) extended the work of KBC13 by testing the consistency of their earlier findings with constraints coming from SN Ia and linear kinematic Sunyaev-Zel\textsc{\char13}dovich effect and showed that local void found by KBC13 could reduce the $H_0$ tension.
    A recent work of \cite{Shanks19} argues that the combination of local inhomogeneity effect together with the re-calibration of Cepheids and local SN Ia distances using the recent parallax measurements from Gaia mission, is sufficient to completely remove the tension [see also \citet{Riess18S,Shanks18} for follow-up discussions].
    \cite{Kenworthy19} (hereafter KSR19) find that any inhomogeneity with contrast$\,>\,20\%$ is strongly inconsistent with SN Ia data, consequently reassuring the ability to measure the distance and Hubble constant with locally calibrated SN to a $1\%$ precision.
    
    The goal of this work is to probe the local matter density using the SN Ia as one of the most reliable low-redshift astrophysical observables, constraining size and amplitude of possible matter inhomogeneity at our local position.
    We investigate the findings of KBC13 and KSR19 by using the complete analytic description for treating the LTB metric in presence of cosmological constant (hereafter $\Lambda$LTB) introduced in \cite{Valkenburg12}.
    Extending the earlier works, we directly fit the $\Lambda$LTB model to two biggest publicly available SN Ia samples - Joint light curve analysis \cite{Betoule14} (hereafter JLA)  and Pantheon compilation \cite{Scolnic17} (hereafter Pan), consisting of 740 and 1048 SN Ia, respectively.
    We also utilise the luminosity density data from KBC13 within the given formalism to assess the (dis)agreement.
    In \cref{sec:th} we review the theoretical description of LTB metric in presence of cosmological constant and proceed by presenting the effect that a local void can have on observable physical quantities in \cref{sec:effs}.
    Constraints on the void contrast and size coming from fitting JLA, Pantheon and KBC13 datasets are reported in \cref{sec:res}, together with the revision of their consistency.
    Additionally, we discuss on the possible effect of local isotropic inhomogeneity on the $H_0$ measurement, and explore the level of anisotropy allowed by the data.
    Summarising our results in \cref{sec:con}, we examine the differences compared to earlier findings by other teams.
    Finally, in the \cref{sec:app} we provide a simplified void model capable of correctly reproducing our results based on complete $\Lambda$LTB description to a good approximation.
    
    Throughout the paper we use the dimensionless variable $h_{70}=H_0/70\,\kmsMpc$ for distance measures; acronym w.r.t. stands for `with respect to'.
    
    \section{Local matter density in $\Lambda$LTB model}
    \label{sec:th}
    The simplest extension of standard homogeneous cosmological model in presence of large matter inhomogeneity is done using the LTB metric \citep{Lemaitre33,Krasinski97} that describes spherically symmetric pressureless cosmic fluid
    \be
    \label{eq:metric}
    \diff s^2=c^2 \diff t^2-\frac{R_r(t,r)^2}{1-k(r)r^2} \diff r^2-R(t,r)^2 \diff\Omega^2.
    \ee
    Here $R(t,r)$ has the units of length and defines all physical distance measures as well as the expansion history in this model.
    The second free function of the metric, $k(r)$ is dimensionless and defines the curvature within the shell of a radius $r$.
    We note that radial coordinate $r$ is also dimensionless, with no physical meaning, and it should be considered only as a flag coordinate of different shells all centred at $r=0$.
    Using the above metric and the Einstein field equations (FE) for pressureless inhomogeneous matter fluid in presence of cosmological constant $\Lambda$, one can derive the generalised Friedmann-Lema\^{i}tre equation as
    \be
    \label{eq:FL}
    \left(\frac{R_t}{R}\right)^2=\frac{2 G M}{R^3}-\frac{c^2 k\,r^2}{R^2}+\frac{c^2\Lambda}{3}.
    \ee
    Here $M(r)$ is an integration constant, whose physical meaning is the total mass enclosed inside a sphere with radial coordinate $r$,
    \be
    \label{eq:Mdeff}
    M(r)=\int_0^r 4\pi R(t,r)^2\rho_m(t,r)R_r(t,r)\,\diff r\,.
    \ee
    Inhomogeneous cosmic matter density profile, $\rho_m(t,r)$, satisfies the conservation law coming from FE, ensuring that integrated mass $M(r)$ remains constant in time.
    In inhomogeneous cosmology we differ two expansion rates, namely transverse and radial:
    \ba
    \label{eq:Hperp}
    H_\perp(t,r)&=\frac{R_t}{R},\\
    \label{eq:Hrad}
    H_{||}(t,r)&=\frac{R_{t r}}{R_r}.
\end{align}
Hence, the one appearing in Friedmann-Lema\^{i}tre \cref{eq:FL} is the transverse expansion rate.
As usual, we denote $t=t_0$ as present age of the Universe and all the physical quantities with subscript "0" represent their values today, while $H_0(r)=H_\perp(t_0,r)$ is used for the transverse expansion rate.

Introducing scale factor, the Friedmann-Lema\^{i}tre equation can be rewritten in a simpler form, although the definition of scale factor $a(t,r)$ is not trivial nor customary as in the homogeneous cosmology.
The value of flag coordinate $r$ is commonly defined through a gauge with some other physical variable. 
The choice of this gauge is arbitrary and has no implications on the model predictions nor on the results of data analysis.
The most often used gauge \citep{Alnes06a,Alfedeel10,Nadathur11} in LTB framework is:
\be
R_0(r)= c t_0 r\quad\Rightarrow\quad r\equiv\frac{R_0(r)}{c t_0}\,.
\ee
However, in this work we use an alternative definition which turns out to be more convenient choice for the latter calculations of elliptical integrals present in the geodesic equations and distance variables.
Following \cite{Valkenburg12}, we define the dimensionless radial coordinate $r$ through the relation with the total mass enclosed in a given shell as
\be
\label{eq:newGauge}
M(r)=\frac{4\pi}{3}C_Mr^3\quad\Longrightarrow\quad r\equiv\sqrt[\leftroot{-1}\uproot{2}\scriptstyle 3]{\frac3{4\pi}\frac{M(r)}{C_M}},
\ee
where $C_M$ is a normalisation constant in units of mass whose value will be chosen at a later stage.
Commonly, in both gauges the scale factor can be defined as
\be
\label{eq:ScaleF}
a(t,r)\equiv\frac{R(t,r)}{c t_0 r}.
\ee
This function clearly converges to the usual scale factor definition and the spatially uniform values in the limit of homogeneous FLRW metric.
Generalised Friedmann-Lema\^{i}tre \cref{eq:FL} can be rewritten in terms of three  dimensionless density parameters normalised to the present time,
\be
\label{eq:FLOm}
H_\perp(t,r)^2={H_0(r)}^2\left(\Omega_m(r)\frac{R_0(r)^3}{R(t,r)^3}+\Omega_k(r)\frac{R_0(r)^2}{R(t,r)^2}+\Omega_\Lambda(r)\right),
\ee
These can be evaluated combining \cref{eq:FL,eq:newGauge,eq:ScaleF,eq:FLOm,eq:Hperp} as,
\ba
\label{eq:Omegam}
\Omega_m&=\frac1{{H_0}^2}\frac{2 G M}{{R_0}^3}=\frac{1}{{H_0}^2}\frac{8 \pi  G C_Mr^3/3}{c^3{t_0}^3{a_0}^3r^3}=\frac{\bar{M}}{{t_0}^2{H_0}^2{a_0}^3}\,,\\
\label{eq:Omegak}
\Omega_k&=-\frac1{{H_0}^2}\frac{c^2k\,r^2}{{R_0}^2}=-\frac{k}{{t_0}^2{H_0}^2{a_0}^2}\,,\\
\label{eq:OmegaL}
\Omega_\Lambda&=\frac{1}{{H_0}^2}\frac{c^2 \Lambda }{3}=\frac{\bar{\Lambda }}{{t_0}^2{H_0}^2}.
\end{align}
Here $\bar{M}=8\pi GC_M/(3c^3t_0)$ and $\bar\Lambda=\frac13\Lambda c^2{t_0}^2$ are dimensionless constants, whereas $k(r)$, $H_0(r)$ and $a_0(r)$ are not uniform in LTB metric.
The introduced dimensionless density parameters are different from one shell to another, but do not depend on time.
Since $\Omega_m(r)$ is proportional to the total mass enclosed in a shell, $M(r)$, it can be represented as the ratio of {\it average} and critical matter densities of each shell.
Starting from \cref{eq:Mdeff} and then using \cref{eq:newGauge,eq:ScaleF} one can define average matter density:
\be
\label{eq:rhoav}
\left<\rho_m\right>(t,r)=\frac{M(r)}{4\pi R(t,r)^3/3}=\frac{C_M}{c^3{t_0}^3}\frac1{a(t,r)^3}\,.
\ee
Then, definition of critical matter density together with \cref{eq:rhoav,eq:Omegam} straightforwardly confirm:
\be
\rho_{\rm crit}\equiv\frac{{3H_\perp}^2}{8\pi G}\quad\Longrightarrow\quad\Omega_m(r)=\frac{\left<\rho_m\right>(t_0,r)}{\rho_{\rm crit}(t_0,r)}\,.
\ee
Friedmann-Lema\^{i}tre \cref{eq:FL} can also be rewritten in dimensionless form using \cref{eq:Omegam,eq:Omegak,eq:OmegaL}
\be
\label{eq:FLsim}
{t_0}^2H_\perp(t,r)^2=\frac{\bar{M}}{a(t,r)^3}-\frac{k(r)}{a(t,r)^2}+\bar{\Lambda}\,.
\ee
In terms of the Birkhoff theorem, \cref{eq:FL,eq:FLOm,eq:FLsim} dictate every shell in $\Lambda$LTB model to evolve as an independent homogeneous FLRW model with a specific curvature $k(r)$, a total mass enclosed $M(r)$ and the same value of cosmological constant $\Lambda$ for all shells, determining its transverse expansion rate $H_\perp(t,r)$.
Each shell is described with relative density parameters $\Omega_i(r)$, present expansion rate $H_0(r)$, and evolving scale factor ${a(t,r)}/{a_0(r)}$.

Cosmological constant, included in the Einstein field equations, is uniform over space and constant in time due to its physical nature, which is also true for the dimensionless constant $\bar\Lambda$ that we introduced.
The density parameter corresponding to the cosmological constant, $\Omega_\Lambda(r)$, can {\it not} be uniform in $\Lambda$LTB model, as this would imply that $\Lambda$ is inhomogeneous and coupled to the cold dark matter density profile.
This approach is reflected in a number of recent studies \cite{Valkenburg12,Tokutake17,Kenworthy19,Boehringer19}, while not fully followed by HB17 who assume cosmic density parameter $\Omega_{\Lambda}$ to be constant throughout the underdense region and at the background.
Similarly, the methods of inverse reconstruction of matter density profile in presence of cosmological constant must be constructed such that FE are satisfied (c.f. \cite{Tokutake16,Wojtak17}).

The gauge choice for the flag coordinate $r$, which is set by \cref{eq:newGauge}, enabled us to rewrite the Friedmann-Lema\^{i}tre \cref{eq:FL} in a more elegant analytic form, i.e. \cref{eq:FLOm}.
The remaining degree of freedom in the scaling constant $C_M$, or equivalently $\bar{M}$, can be used to normalise the scale factor $a(t,r)$.
We set $\bar{M}$ such that $a_*=a\left(t_0,r_*\right)=1$ for a chosen shell $r=r_*$ and at the present time $t_0$.
For example, one can choose $r_*=0$ to normalise the scale factor at the observer's position or $r_*=+\infty$ to normalise at the background, instead.
We follow the normalisation of scale factor on the background, setting the numerical value of $\bar{M}$ such that $a_*=a\left(t_0,r_*=+\infty\right)=1$ for the background shells.
In particular, normalisation gives the physical meaning to the scaling constant, i.e. \cref{eq:rhoav} yields
\be
\label{eq:rhobg}
\rho_m^{\rm bg}(t)=\frac{\rho_m^{\rm bg}(t_0)}{a^{\rm bg}(t)^3}\quad\Longrightarrow\quad \rho_m^{\rm bg}(t_0)={C_M}{c^3{t_0}^3}\,.
\ee
Therefore, normalising far from the inhomogeneity, which is for any large enough $r_*$, implies that background scale factor and \cref{eq:FLsim} converge to the usual forms that we have in homogeneous FLRW model.
It is easy to see that at present time $t=t_0$ the background quantities satisfy
\be
t_0H_0^{\rm bg}=\sqrt{\bar{M}-k^{\rm bg}+\bar{\Lambda}}\,.
\ee
Partial differential \cref{eq:FLsim} can be used to express the derivative $\partial t/\partial a$ and integrate from the Big Bang up to scale $a$:
\be
\label{eq:tInt}
\frac{t(a,r)-t_{BB}(r)}{t_0}=\int _0^a\frac{\sqrt{\alpha }}{\sqrt{\bar{\Lambda}\,\alpha^3-k(r)\,\alpha+\bar{M}}}\diff \alpha\,.
\ee
All variables on the right hand side are dimensionless and this elliptical integral is analytically solvable (for details see \cite{Valkenburg12}).
The Big Bang time $t_{BB}(r)$ is an integration constant that may vary in space, affecting the age difference between the shells.
Nevertheless, one does not expect large variation of $t_{BB}(r)$ in the case of a small local inhomogeneity.
Non-uniform Big Bang time is also related to instabilities of matter perturbations \citep{Zibin08}.
Hence, in this work we consider only the models with homogeneous Big Bang time $t_{BB}(r)\equiv{\rm const}=0$.

Integrating \cref{eq:tInt} on the background ($r=r_*$) from the Big Bang $a=0$ up to today $a(t=t_0,r_*)=1$ we have
\be
1=\frac{t(a=1,r_*)}{t_0}=\int _0^1\frac{\sqrt{\alpha }}{\sqrt{\bar{\Lambda}\,\alpha^3-k^{\rm bg}\,\alpha+\bar{M}}}\diff \alpha,
\ee
which is the numerical equation defining a value of $\bar{M}$, necessary for the normalisation.
At any other point of interest \cref{eq:tInt} can be seen as the solution for $t(a,r)$ or, after invertion, as $a(t,r)$.
Hence, calculating numerically the scale factor at any spatial position and time gives us the angular diameter distance $R(t,r)$ and other physical observables of interest.

\section{Effect on astrophysical observables}
\label{sec:effs}
The high-redshift background observables  are not affected by the local isotropic inhomogeneity.
Nevertheless, the analysis of local observables can yield different constraints on cosmological parameters from those obtained with FLRW models.
Examples presented in this section form the basis for analysing observable data, but do not immediately represent the effect on cosmological inferences.
We use a specific analytic form of the curvature, parametrised with ${k^{\rm bg},k^{\rm loc},r_0,\Delta r}$ and given as Garc\'ia-Bellido-Haugb{\o}lle (GBH) profile \citep{Garcia08,February10}:
\be
k(r)=k^{\rm bg} - \left(k^{\rm bg} - k^{\rm loc}\right) \frac{1 - \tanh\left(\frac{r - r_0}{2 \Delta r}\right)}{
1 + \tanh\left(\frac{r_0}{2 \Delta r}\right)}\,.
\ee
Our focus is the case of $\Lambda$CDM background, i.e. $k^{\rm bg}\equiv0$, which we call $\Lambda$LTB model.
The analytic form of $k(r)$ together with the value of dimensionless cosmological constant $\bar{\Lambda }$ define the evolution of expansion and all physical quantities in LTB metric.
In the limit of $\Delta r \rightarrow 0$, a simpler top-hat profile (TH) is recovered,
\ba
k(r)&=k^{\rm loc}&&\quad{\rm for}\,\,r<r_0\,,\hspace{12em}\nonumber\\
k(r)&=k^{\rm bg}&&\quad{\rm for}\,\,r\geq r_0\,.\hspace{12em}
\end{align}
While GBH form describes a smooth matter density profile, TH profile has one parameter less and is trivial to use for data fitting as two joint homogeneous models.
Care must be given to relation between the physical quantities in two homogeneous regions.
\subsection{Reconstruction of geodesic}
The LTB metric describes an inhomogeneous dust cloud with spherical symmetry, {indicated}  by the angular term in \cref{eq:metric}.
Hence, the observer located at the centre of this symmetry has an isotropic view.
The assumption of isotropy gives us only the first approximation for treating the large cosmological inhomogeneities.
To quantify the overall effect of large local void on cosmological inferences, we limit ourselves to the on-centre model that considers an observer located at the centre of symmetry in the local matter inhomogeneity.
The assumption of this very special position can be avoided, although the geodesic equations for an off-centre observer become increasingly complex, affecting the feasibility of data analysis.
We find the on-centre model as adequate for presenting all the relevant points in this work, but we extend the discussion about observer's spatial position later on.

Geodesic equations can be derived following \cite{Celerier00}:
\ba
\label{eq:drdz}
\frac{\diff r}{\diff z}&=\frac1{1+z}\frac{c\sqrt{1-k\,r^2}}{R_{t r}}\,,\\
\label{eq:dtdz}
\frac{\diff t}{\diff z}&=-\frac1{1+z}\frac{R_r}{R_{t r}}\,.
\end{align}
where $z$ is the cosmological redshift.
This system of differential equations must be solved numerically along the geodesic with help of \cref{eq:FLsim,eq:ScaleF,eq:Hperp,eq:tInt}, in the entire redshift range.
In order to derive distance-redshift relations one needs to numerically reconstruct the function $R(z)$, or equivalently $a(z)$, which can be obtained by numerical inversion of \cref{eq:tInt} at every $z$ along the geodesic \citep{Valkenburg12}.
However, we find it easier and computationally faster to use the geodesic equations and all physical variables of interest rewritten with $(a,r)$ as independent fundamental coordinates, after the change of variables from $(t,r)$
\be
\frac{\diff a}{\diff z}=a_t\frac{\diff t}{\diff z}+a_r\frac{\diff r}{\diff z}.
\ee

Adding the assumption of homogeneity, geodesic equations reduce to the standard form; using \cref{eq:dtdz,eq:ScaleF} we have:
\be
\label{eq:dadzH}
\frac{\diff t}{\diff z}=-\frac1{1+z}\frac{a+r\,a_r}{a_t+r\,a_{t r}}=-\frac1{1+z}\frac a{a_t}\quad\Rightarrow\quad\frac{\diff a}{a}=-\frac{\diff z}{1+z}\,.
\ee
We note that this relation is valid on all shells where spatial derivative is null, i.e. in every homogeneous subregion along the matter density profile.
\subsection{Density contrasts}
\begin{figure}
\includegraphics[width=.48\textwidth]{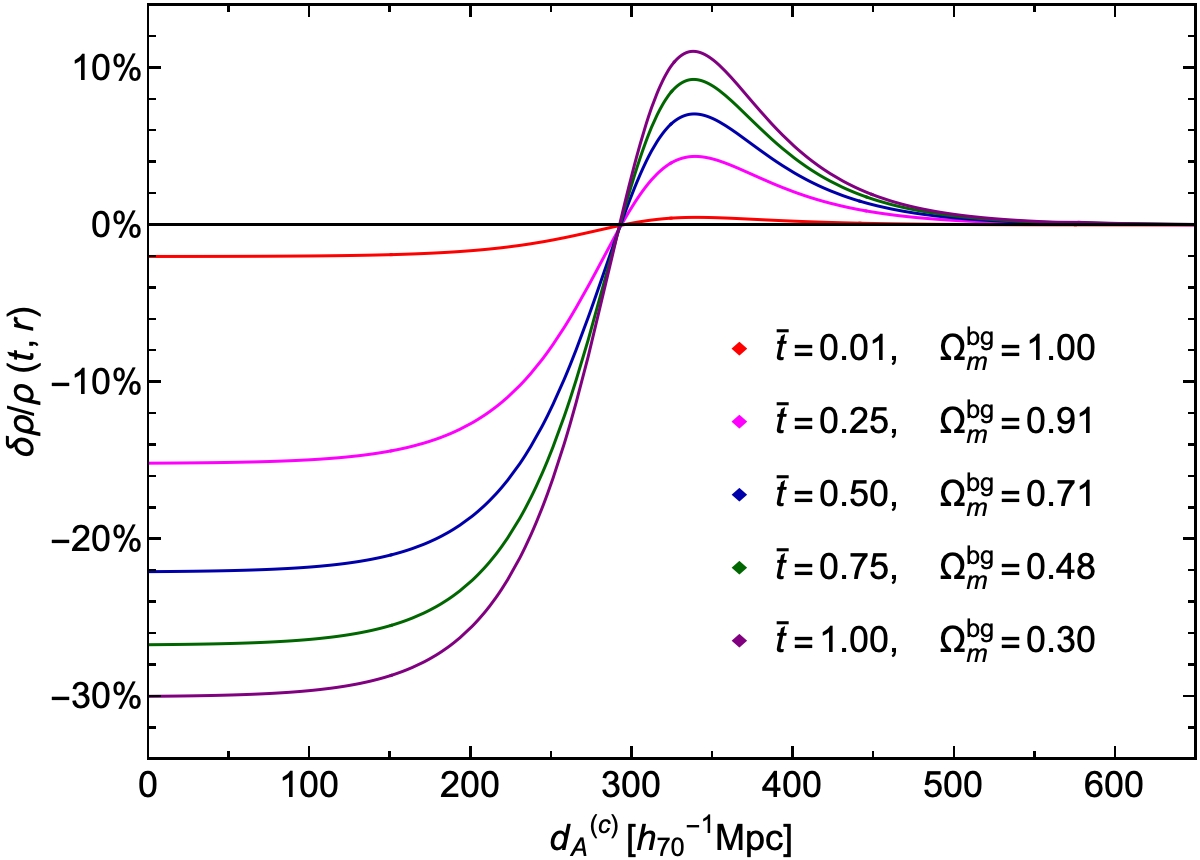}
\caption{The evolution of matter density contrast $\delta\rho/\rho$ is shown for $\Lambda$LTB model with present local contrast $\delta\rho_0=-30\%$ and redshift size $z_{\rm size} = 0.07$. The profile is shown in comoving angular diameter distance, at time lapses $\bar{t}=t/t_0$ running from 0.01 up to 1.00 (present time).}
\label{fig:contrast}
\end{figure}
It is useful to define the relative contrasts for quantities of interest, such as physical matter density, dimensionless density parameter $\Omega_m$, and expansion rates.
The contrast can be defined at any time and any spatial point inside inhomogeneous region. 
Consequently, for physical matter density we have
\be
\label{eq:deltarho}
\frac{\delta\rho_m}{\rho_m}=\frac{\rho_m(t,r)}{\rho_m^{\rm bg}(t)}-1
\ee
The evolution of matter density contrast is shown in \Cref{fig:contrast} for a $\Lambda$LTB void model with specific parameters determining cosmological constant $\bar\Lambda$, central curvature $k^{\rm loc}$, size and steep transition zone, given with $r_0$ and $\Delta r$.
As shown in this case, the physical matter density $\rho_m$ will have compensated radial profile in the case of small $\Delta r$ and/or large central contrast.
Since we use the comoving radial distance, the physical expansion of void size can not seen on \Cref{fig:contrast}.

Most often we consider the present ($t=t_0$) contrast between observer's position ($r=0$) and background ($r=+\inf$) and we denote it with subscript 0.
The contrast of dimesionless matter density parameter $\Omega_m(r)$ and both expansion rates are defined in the same manner as for physical matter density in \cref{eq:deltarho}.
We will focus on present central contrasts which can be derived using \cref{eq:rhobg,eq:Mdeff,eq:newGauge,eq:ScaleF,eq:Omegam}:
\ba
\label{eq:drho}
{\delta\rho_0\over\rho_0}=&\frac{\rho_m^{\rm loc}(t_0)}{\rho_m^{\rm bg}(t_0)}-1=\left(\frac1{a_0^{\rm loc}}\right)^3-1\,,\\
\label{eq:dOm}
{\delta\Omega_0\over\Omega_0}=&\frac{\Omega_m^{\rm loc}}{\Omega_m^{\rm bg}}-1=\left(\frac{H_0^{\rm bg}}{H_0^{\rm loc}}\right)^2\left(\frac1{a_0^{\rm loc}}\right)^3-1\,.
\end{align}
Notably, these two contrast are different due to the induced spatial inhomogeneity of the expansion rates,
\be
\label{eq:dH}
{\delta H_0\over H_0}=\frac{H_0^{\rm loc}-H_0^{\rm bg}}{H_0^{\rm bg}}\,.
\ee
In conclusion, relation $\left|\delta\rho_0/\rho_0\right|<\left|\delta\Omega_0/\Omega_0\right|$ holds both in the case of local underdensity as well as overdensity.
Writing the equivalent of \cref{eq:Mdeff} at $a(t=t_0,r=0)=a_0^{\rm loc}$ and after some simple calculations, one can derive analytic expressions that relate these three contrasts.

We would like to note that the current $\Lambda$LTB modelling can be equivalently parametrised with $\Omega_m^{\rm bg}$, $\delta\rho_0/\rho_0$, $z_{\rm size}$ and $\Delta z$, which are derived from $\bar\Lambda$, $k^{\rm loc}$, $r_0$ and $\Delta r$.
Approximate series expansion formulae, valid for $\Lambda$CDM background, are provided in \cref{sec:app}.
\subsection{Scale factor}
The radial dependence of scale factor in LTB models can be observed only along the geodesic as $a(z)=a(t(z),\,r(z))$ (see \Cref{fig:Hscale}).
Moreover, modelling $a(z)$ can be used a posteriori to inversely reconstruct the matter density profile \citep{Wojtak17}.

Inside homogeneous region close to the centre of GBH profile and at the background, scale factor satisfies \cref{eq:dadzH} and, hence:
\ba
a(z)=&{a_0^{\rm loc}\over1+z}\qquad\text{for }z<<z_{\rm size}\,,\\
a(z)=&{1\over1+z}\qquad\text{for }z>>z_{\rm size}\,.
\end{align}
In fact, redshift size of an inhomogeneous matter profile can be defined through the equality
\be
2\,(1+z_{\rm size})\,a(z_{\rm size})=a_0^{\rm loc}+1\,.
\ee
\subsection{Local expansion rate}
\begin{figure}
\includegraphics[width=.48\textwidth]{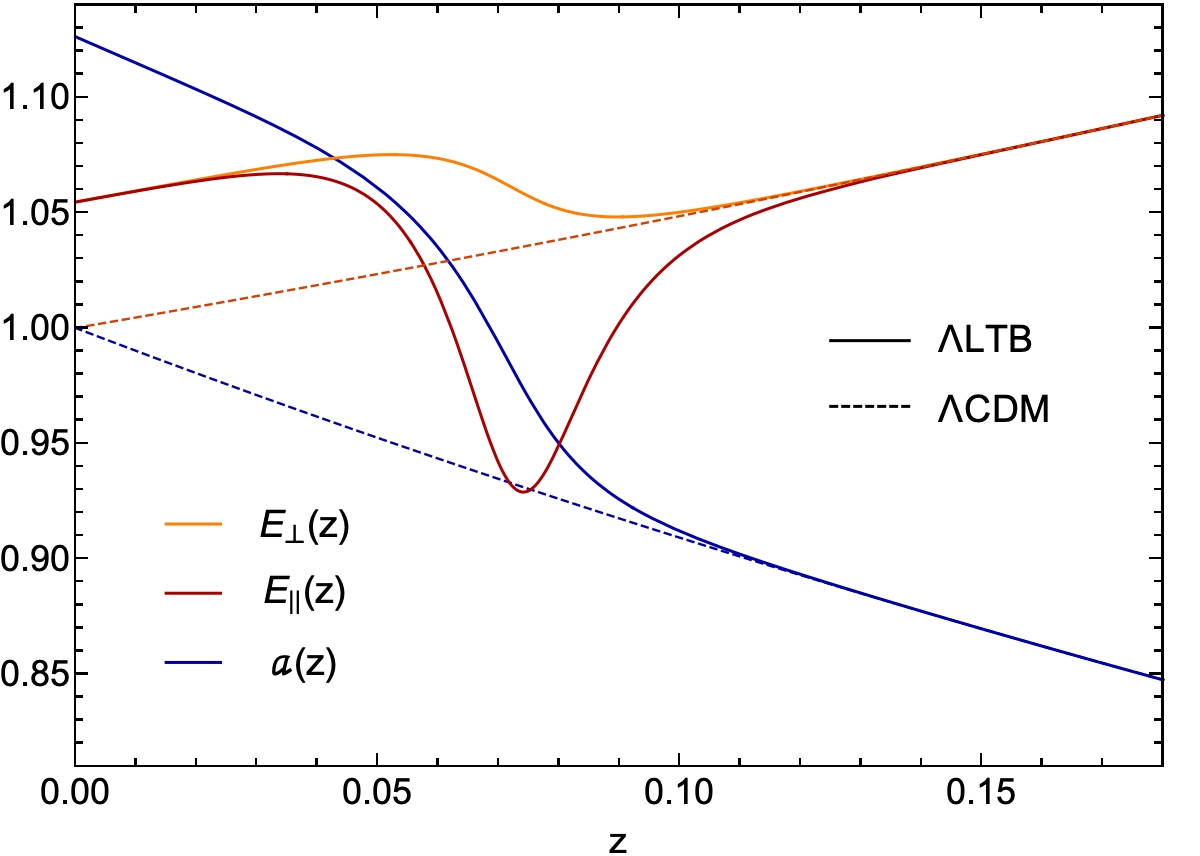}
\caption{Redshift dependence of the cosmic expansion rates (red for radial and orange for transverse) and the scale factor (blue) are shown for $\Lambda$LTB model with contrast $\delta\rho_0=-30\%$ and redshift size $z_{\rm size} = 0.07$ (solid lines) in comparison with the $\Lambda$CDM model (dashed lines).}
\label{fig:Hscale}
\end{figure}
As explained in previous section, LTB metric is characterised with two expansion rates, shown in \Cref{fig:Hscale}. 
The transverse expansion rate given by Friedmann-Lema\^{i}tre \cref{eq:FLOm} can be used together with density parameters $\Omega_i$ to construct the equivalent FLRW model of each shell.
On the contrary, radial expansion rate is related to the spatial derivative along the line of sight (cf. \cref{eq:Hrad}).
Certainly, all physical quantities defined in $\Lambda$LTB model converge to their $\Lambda$CDM form at high redshift limit as a consequence of the Birkhoff theorem for the local spherical inhomogeneity (see \Cref{fig:Hscale}).

The intercept of SN Ia $m(z)$ relation is determined by the absolute magnitude of SN and the local expansion rate.
Calibrated SN Ia are able to measure the $H_0$ only after fixing the slope of distance-redshift relation as presented in \cite{Riess16}.
However, if the inhomogeneity extends up to $z$\,$\sim$\,$0.1$, not accounting for its effect properly can represent an important systematic in $H_0$ measurement.
\section{Results and Discussion}
\label{sec:res}
In this section we present our main findings in the analysis of LD data from KBC13 and the SN Ia data from JLA and Pantheon\footnote{We use the latest release of Pantheon dataset from Nov 2018, while verifying that using the SN Ia redshifts published earlier has minimal to no effect on the likelihood in all fits performed here (see also \cite{Rameez19})} samples \citep{Betoule14,Scolnic17}.
The likelihoods for SN Ia datasets are implemented as in their respective releases.
Results are presented for the derived parameters: relative matter density at the background $\Omega_m^{\rm bg}$, central contrast $\delta\Omega_0/\Omega_0$, inhomogeneity size $z_{\rm size}$ and density profile shape $\Delta z$.
We use a prior of $\Omega_m^{\rm bg}=0.3153\pm0.0073$ taken from CMB TT,TE,EE+lowE+lensing data analysis for the $\Lambda$CDM background \citep{Planck18}, which appropriately aids low-redshift SNe to constrain the local matter density profile (hereafter indicated as JLA+P and Pan+P).
Other parameters are sampled from wide flat prior ranges: $-0.4\leq\delta\Omega_0/\Omega_0\leq0.2$, $z_{\rm size}\leq0.13$ ($\approx$\,0.5Gpc in comoving distance) and $\Delta z\leq0.6\,z_{\rm size}$, for which we verify a posteriori that relaxing the priors does not affect our results.

As elaborated in \cref{sec:effs}, the large local void is expected to affect the magnitude-redshift relation of SN Ia inside the void, due to the higher transverse expansion rate and lower values of relative density parameters $\Omega_m(z)$ and $\Omega_\Lambda(z)$ w.r.t. $\Lambda$CDM constraints.
We start our analysis by testing for redshift dependence of SN Ia magnitude parameter
\be
{\cal M}=M_B+5\rm{Log}_{10}(c/H_0).
\ee
To this end, we use Pantheon SN Ia whose apparent magnitudes have already been corrected with a cosmology-independent method which marginalises the effect of all magnitude correction parameters and leaves only ${\cal M}$ as a free parameter, besides the cosmological model (see \cite{Scolnic17}).
\begin{figure}
\includegraphics[width=.46\textwidth]{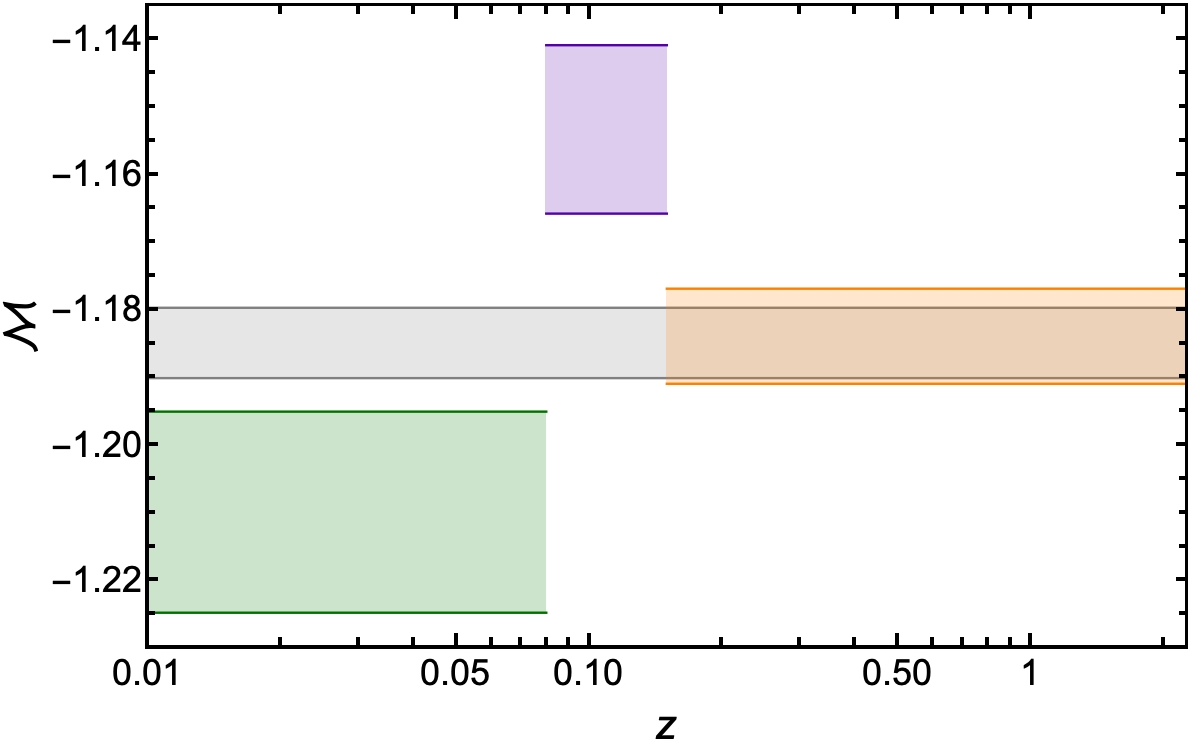}
\caption{$1\sigma$ confidence stripes of SN Ia magnitude parameter $\cal M$ obtained with standard (grey) and binned: $z\leq0.08$ (green), $0.08<z\leq0.15$ (violet) and $z>0.15$ (orange) fits to Pan+P data.}
\label{fig:itMPan}
\end{figure}
Full Pantheon dataset, along with the CMB prior, is fitted to standard $\Lambda$CDM model, allowing for three binned values of $\cal M$ in redshift ranges $z\leq0.08$ (green), $0.08<z\leq0.15$ (purple) and $z>0.15$ (orange), with 194, 103 and 751 SNe, respectively.
The resulting $1\sigma$ confidence interval for ${\cal M}^{\rm bg}=-1.184\pm0.007$ is in excellent agreement with ${\cal M}^{\rm tot}=-1.185\pm0.005$ from the standard non binned fit shown as grey stripe in \Cref{fig:itMPan}.
The visible inconsistency in the first two bins, ${\cal M}^1=-1.210\pm0.015$, ${\cal M}^2=-1.153\pm0.012$ might be hinting to possible local geometry effects.
Fitting the Pantheon SN Ia in the redshift range $0.0233<z<0.15$ to the FLRW series expansion (SE) formula for luminosity distance with $q_0=-0.55$, as presented in R16, we obtain ${\cal M}({\rm SE})=-1.182\pm0.010$, which is also different from values obtained in the first two bins.
Various works in the literature, including R16 and their previous articles, introduce a parameter $a_B$ that can be derived from the intercept magnitude $\cal M$ as
\be
a_B=-0.2{\cal M} -5 + \log_{10}c\,,
\ee
where $c$ is the speed of light in km/s.
 
Results from the simple binned analysis already provide a strong motivation to study a more physical, inhomogeneous cosmological model, characterised by varying $H_0(r)$ and matter density profile $\Omega_m(r)$.
Although an extended analysis of the $\cal M$ evolution at higher redshifts could have important cosmological implications (see e.g.  \cite{Tutusaus19} for effect on cosmic acceleration), here we looked for possible local variation in $\cal M$ as a hint of local matter inhomogeneity, while our background cosmology is assumed to be flat $\Lambda$CDM.
\begin{figure}
\includegraphics[width=.48\textwidth]{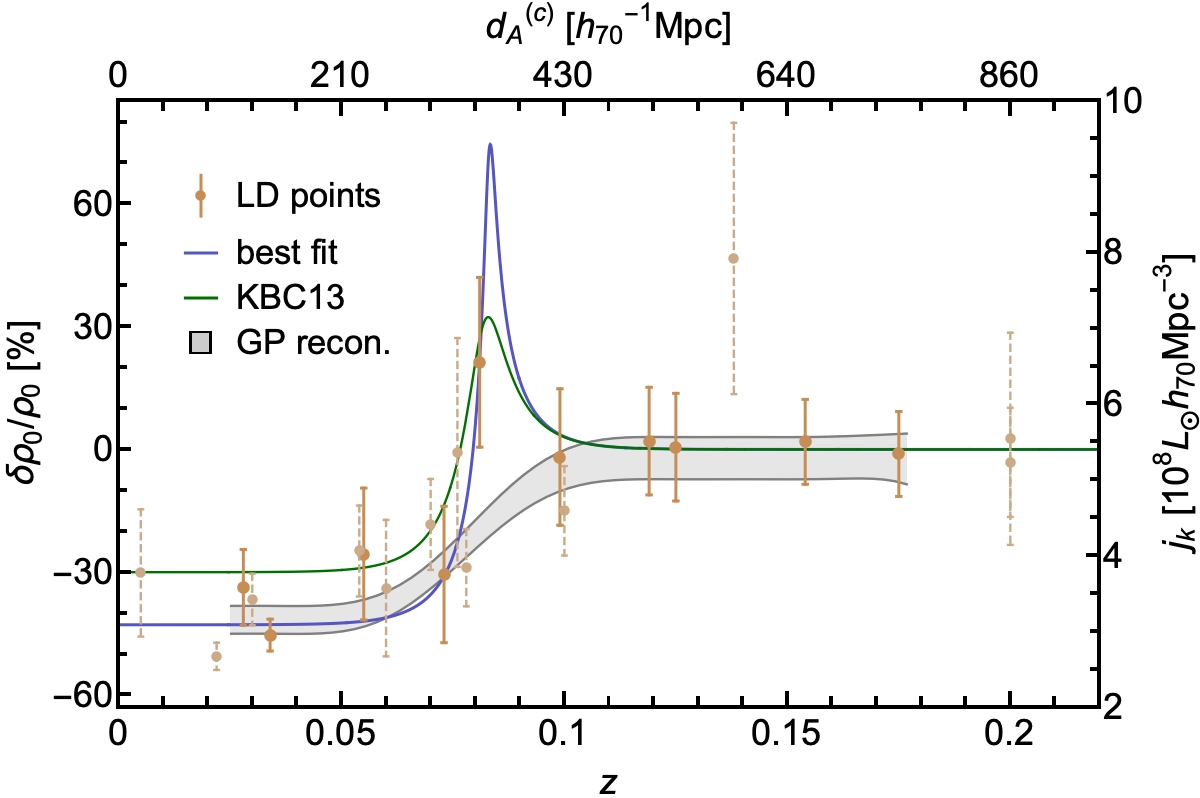}
\caption{Luminosity density data from KBC13: solid orange data points are their original data, while dashed points are from additional surveys. The grey area is a $1\sigma$ reconstruction zone obtained using Gaussian process. The blue curve is our best-fit GBH profile to 10 LD points, which shows $\delta\rho_0/\rho_0=-43\%$ contrast. The green curve has a weaker contrast $\delta\rho_0/\rho_0=-30\%$ suggested by KBC13.}
\label{fig:GBHKBC}
\end{figure}

{\renewcommand{\arraystretch}{1.75}%
\begin{table*}
\centering
\caption{Primary results from the $\Lambda$LTB analysis performed to: luminosity density using all 10 original data points presented in KBC13 (LD) and after removal of $z=0.034$ data point from 2MASS galaxy catalogue (LD$^*$); 740 SN Ia from JLA dataset alone (JLA) and with Planck constraint on $\Omega_m^{\rm bg}=0.3153\pm0.0073$ (JLA+P); 1048 SN Ia from Pantheon dataset (Pan) and with Planck constraint (Pan+P); 237 low-redshift Pantheon SN Ia from the range $0.0233<z<0.15$ with the $\Omega_m^{\rm bg}=0.3$ constraint (Pan$^*$). The best-fit values for parameters of interest (noted as $^{b.f.}$) are provided together with confidence interval between 16th and 84th percentiles (noted as $^{c.i.}$) of their respective 1D marginalised likelihoods. Three contrast presented here are defined by \Cref{eq:drho,eq:dOm,eq:dH}. The comparison to $\Lambda$CDM model is given using the relative $\Delta$AIC values.}
\begin{tabular}{l c c c c c c c c c c}
\hline
data & $z_{\rm size}^{b.f.}$ & $z_{\rm size}^{c.i.}$ & ${\delta\Omega_0\over\Omega_0}^{b.f.}$ [\%] & ${\delta\Omega_0\over\Omega_0}^{c.i.}$ [\%]& ${\delta\rho_0\over\rho_0}^{b.f.}$ [\%]& ${\delta\rho_0\over\rho_0}^{c.i.}$ [\%]& ${\delta H_0\over H_0}^{b.f.}$ [\%]& ${\delta H_0\over H_0}^{c.i.}$ [\%]  & $\chi^2$ & $\Delta$AIC$_{\Lambda{\rm CDM}}$\\
\hline
LD  & $0.082$ & $0.079^{+0.012}_{-0.012}$  & $-51.1$ & $-51.9^{+6.3}_{-6.3}$ & $-42.9$ & $-43.8^{+6.0}_{-6.1}$ & 8.1 & $8.2^{+1.3}_{-1.2}$ & 2.62 & \slashbox{}{}\\
LD$^*$  & $0.082$  & $0.075^{+0.015}_{-0.015}$  & $-39.7$ & $-39.4^{+10.3}_{-10.3}$ & $-32.4$ & $-32.1^{+9.0}_{-9.5}$ & 5.9 & $5.9^{+1.9}_{-1.7}$ & 0.29 & \slashbox{}{}\\
JLA  & $0.025$  & $0.039^{+0.062}_{-0.018}$  & $-19.5$ & $-9.9^{+17.3}_{-13.9}$ & $-15.1$ & $-7.5^{+12.9}_{-11.0}$ & 2.7 & $1.3^{+2.0}_{-2.3}$ & $678.30$ & $1.35$\\
JLA+P  & $0.025$  & $0.032^{+0.056}_{-0.011}$  & $-19.6$ & $-12.5^{+13.8}_{-12.4}$ & $-15.2$ & $-9.5^{+10.4}_{-9.9}$ & 2.7 & $1.7^{+1.9}_{-1.9}$ & $678.34$  & $1.07$ \\
Pan  & $0.075$  & $0.070^{+0.023}_{-0.031}$  & $-16.2$ & $-9.8^{+14.0}_{-8.9}$ & $-12.4$ & $-7.4^{+10.5}_{-7.0}$ & 2.2 & $1.3_{-1.9}^{+1.3}$ & $1020.72$  & $0.67$ \\
Pan+P  & $0.075$  & $0.068^{+0.021}_{-0.030}$  & $-14.4$ & $-10.5^{+9.3}_{-7.4}$ & $-11.0$ & $-7.9^{+7.0}_{-5.8}$ & 2.0 & $1.4_{-1.3}^{+1.1}$ & $1021.01$  & $0.40$ \\
Pan$^*$  & $0.075$ & $0.075^{+0.013}_{-0.022}$  & $-23.1$ & $-21.5^{+8.9}_{-9.5}$ & $-18.1$ & $-16.8^{+7.1}_{-7.9}$ & 3.2 & $3.0^{+1.5}_{-1.3}$ & $225.63$  & $-3.30$ \\
\hline
\end{tabular}
\label{tab:1}
\end{table*}}

\subsection{$\Lambda$LTB analysis with LD data:}
\label{ssec:resLD}
KBC13 summarises 22 LD data points obtained from several surveys, of which 10 are their original data points. 
Due to the unknown correlations that could arise from the overlap between different redshift surveys we opt to utilise only these 10 data points to constrain local matter density profile. 
They were obtained using data from 35,000 galaxies divided in 10 redshift bins in the range $0.005 < z < 0.20$. 
Using LD points, KBC13 report the existence of a large local void with a luminosity density contrast of about $-30\%$ in the inner region, which is expected to be proportional to the contrast of total matter density $\delta\rho_0/\rho_0$ (see \Cref{fig:GBHKBC}).
We initially perform a simple reconstruction of the density profile using the Gaussian process -- a model-independent technique for data analysis (see e.g. \cite{Holsclaw10, Seikel12a, Haridasu18_GP}).
The reconstructed $1\sigma$ region shown in grey on \Cref{fig:GBHKBC} is obtained by imposing homogeneity both inside and outside the expected void size.
As such, it is allowed to predict no evidence of a void, but our resulting posterior of the reconstructed region shows definite indications of the local underdensity with $\delta\rho_0/\rho_0\sim-40\%$.
We proceed by fitting LD data points to the full $\Lambda$LTB model and confirm that the conservative density contrast of $\delta\rho_0/\rho_0\sim-30\%$, suggested by KBC13 and later utilised in HB17 and KSR19, appears to be an underestimate w.r.t. our best-fit shown in \Cref{tab:1}.

Due to the small amount of data, we test our findings by performing a leave one out (LOO) analysis, implemented as random removal of each of the 10 points from the fit.
It clearly shows the stability of the estimated redshift size, while the value of the profile contrast $\delta\rho_0/\rho_0\sim-44.0\pm 6.0$ is found to be dominated by one stringent measurement at $z = 0.034$.
Eliminating this data point coming from 2MASS survey, the remaining 9 points from UKIDSS and GAMA surveys estimate the contrast to be $\delta\rho_0/\rho_0\sim-32.0\pm 9.5$, in consistence with the conservative value suggested by KBC13.
A direct comparison of our best-fit profile (blue) and the profile with $\delta\rho_0/\rho_0=-30\%$ (green) in \Cref{fig:GBHKBC} clearly shows the importance of 2MASS data point at $z = 0.034$.
Nevertheless, the induced contrast of fractional matter density $\delta\Omega_0/\Omega_0$ coming from the fit to 9 points (quoted as LD$^*$ in \Cref{tab:1}) is still higher than the one used in previous works where the difference between $\delta\Omega/\Omega$ and $\delta\rho/\rho$ contrasts is not considered (c.f. KSR19, HB17, \cite{Shanks19}).

The primary inferences of this analysis are verified to remain unaltered changing the assumed analytic form of matter density profile. 
Taking the result in \Cref{tab:1} at face value, would indicate an extreme tension with a homogeneous matter density scenario.
While the standard model of cosmology expects variation of the local matter density and velocity fields, a more pressing problem arising from this result is the size of the estimated inhomogeneity.
Based on $\Lambda$CDM matter power spectrum constrained by CMB, density perturbations of size $z_{\rm size}\approx0.08$ are expected to have the variance of $\delta\rho_0/\rho_0\approx1.5-2\%$ (corresponds to $\delta\Omega_0/\Omega_0\approx2-3\%$) at $68\%$ confidence level \citep{Marra13,Camarena18}.
Hence, this expectation is at $\sim$\,$8\sigma$ tension with the result of the fit to complete LD data, while the exclusion of 2MASS data point at $z =0.034$ relaxes deviation to a however large $\sim$\,$4\sigma$ tension.

The best-fit $\Lambda$LTB (blue) curve in \Cref{fig:GBHKBC} is closely following the predicted model-independent reconstruction (grey), except for the short overdense region, which is an expected compensation of a large and steep underdense matter profile.
In fact, the data point at $z=0.081$ might be a signature of the same. 
Although not statistically significant to draw conclusions for an overdensity, it aids to constrain the void size.
The specific interplay between the model description and the fact that only 10 data points are available for constraining 4 profile parameters, results in over-fitting difficulties. 
For this reason we perform a Gaussian approximation of the posterior likelihood distribution before estimating the confidence regions presented in \Cref{fig:GBHPanKBC}, which does not alter our primary inferences.

Given the lack of consistency in reported results based on LD data, limited sky coverage of the data, reports about angular variation (WS14), and possible effects of binning we also deem it important to estimate the level of (dis)agreement with the SN Ia data.

\subsection{$\Lambda$LTB analysis with SN Ia datasets:}
\label{ssec:resSN}
Probing for signatures of large local inhomogeneity (underdense or overdense) in the SN Ia datasets, we fit the local matter density profile simultaneously with the background cosmology ($\Omega_m^{\rm bg}$) which is, nevertheless, dominantly guided by the Planck CMB constraint.
Since the variation of $H_0(r)$ is modelled, in $\Lambda$LTB we define the SN Ia intercept magnitude parameter using background expansion rate
\be
\label{eq:itMLTB}
{\cal M}=M_B+5\rm{Log}_{10}(c/H_0^{\rm bg}).
\ee
In the case of JLA dataset, fitting $\chi^2$ also depends on SN Ia magnitude correction parameters for lightcurve stretch, colour and host galaxy mass, which are later marginalised as nuisance parameters \citep{Betoule14}.

\begin{figure}
\includegraphics[width=.48\textwidth]{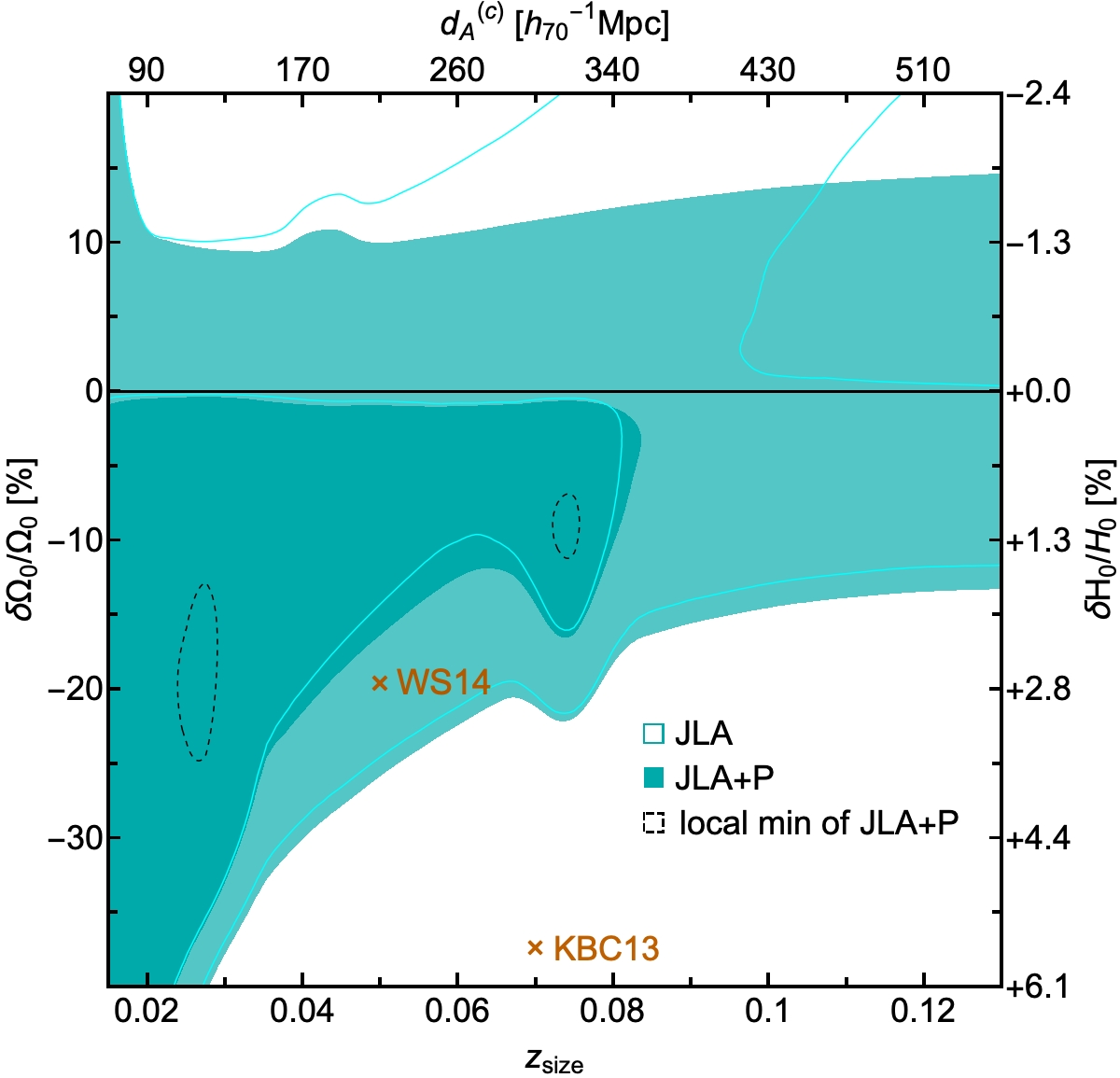}
\caption{Constraints on the local matter density profile from JLA (solid curves) and JLA+P (coloured regions) are shown at $1\sigma$ and $2\sigma$ confidence level. The positions of the only two local $\chi^2$ minima are emphasised with the dashed lines. The isotropic voids proposed by WS14 and KBC13 are marked with orange crosses.}
\label{fig:GBHJLA}
\end{figure}

\begin{figure}
\includegraphics[width=.48\textwidth]{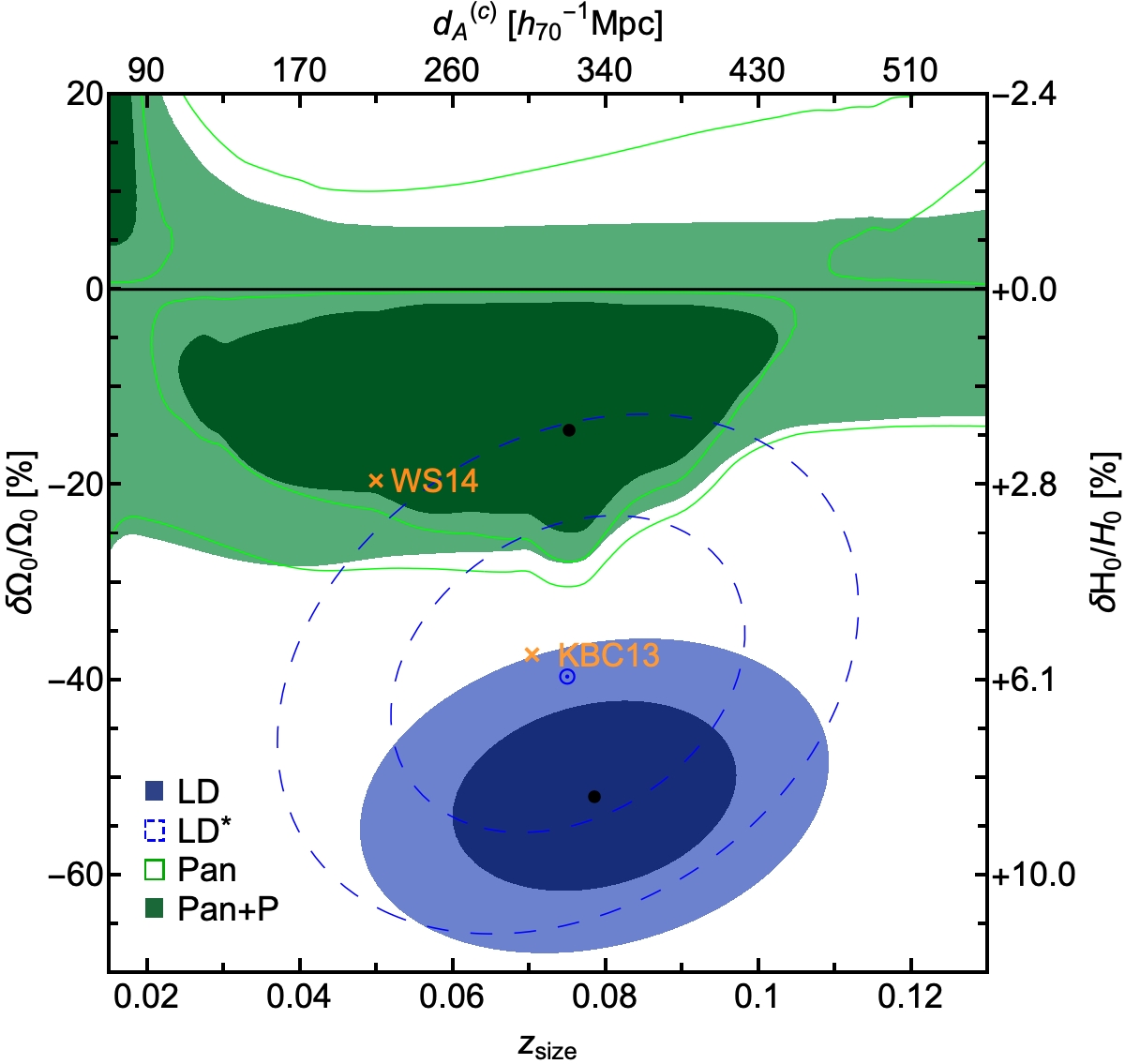}
\caption{$1\sigma$ and $2\sigma$ confidence contours of local matter density profile parameters are shown for Pan+P (green regions), Pantheon (light green solid curves), full LD data from KBC13 (blue regions) and reduced LD$^*$ data (dashed blue curves). The isotropic voids proposed by WS14 and KBC13 are marked with orange crosses. Contrasts presented on vertical axes are defined by \Cref{eq:dOm,eq:dH}.}
\label{fig:GBHPanKBC}
\end{figure}

The SN Ia constraints on local matter density profile presented in the form of confidence regions in \Cref{fig:GBHPanKBC,fig:GBHJLA} and \Cref{tab:1} show high consistency with homogeneous $\Lambda$CDM fit, characterised by $\delta\Omega_0/\Omega_0=0$.
The analysis of JLA dataset yields two distinguished local $\chi^2$ minima, emphasised with dashed lines in \Cref{fig:GBHJLA}. The underdensity reported by WS14, with the size of $\approx$\,$215{h_{70}}^{-1}$Mpc and the average contrast $\delta\rho_0/\rho_0=-15\pm3\%$ (corresponds to $\delta\Omega_0/\Omega_0=-19\pm4\%$) is in between the two minima we find for JLA dataset.
Interestingly, the second local minimum specifies a void with the physical size in line with the findings of KBC13, ${d_A}^{(c)}=300{h_{70}}^{-1}$Mpc, but with a much weaker contrast.
Indeed, the large local void reported by KBC13 is rejected at $\gtrsim$\,$4\sigma$ confidence by JLA dataset.
However, our analysis of reduced LD$^*$ data points gives less stringent constraint on the void contrast which is no longer in tension with our JLA result (cf. \Cref{tab:1}).

Even though the analysis of Pantheon dataset does not recover the primary minimum obtained for JLA supernovae, the resulting confidence regions are of high similarity.
Our Pantheon and Pan+P constraints for $\Lambda$LTB model shown in \Cref{fig:GBHPanKBC} include inside $1\sigma$ confidence region voids with contrast $\delta\Omega_0/\Omega_0\approx-20\%$, such as the representative void of WS14.
The constraints from LD data of KBC13 (blue shaded ellipses in \Cref{fig:GBHPanKBC}) are pointing to much deeper void, although the redshift size of the two best-fits are aligned.
In fact, the estimated contrasts from Pan+P and LD data are at $\sim$\,$4\sigma$ tension (see \Cref{tab:1}).
The nature of inferences resulting from LD data, their disagreement with the expected cosmic variance and the SN Ia constraints, further strengthens our motive to perform LOO analysis to test this data sample.
LOO analysis led us to consider the fit to the reduced LD$^*$ dataset, which is shown with blue dashed contours in \Cref{fig:GBHPanKBC}.
As mentioned earlier, the LD$^*$ constraint on void size remains unaltered, while the estimated void contrast shifts towards a shallow inhomogeneity that has {\it no tension} with the Pantheon results. 

Our $\Lambda$LTB inferences coming from Pantheon dataset are contrasting the recent result of KSR19 which reports a very high significance of $\sim$\,$5\sigma$ against any large local void with contrast $\left|\delta\rho_0/\rho_0\right|>20\%$.
The higher amount of SN Ia data points that they used, 1295 compared to 1048 available in Pantheon, are additionally populating the low-redshift range and are expected to improve the error bars of profile parameters by $\gtrsim$\,$50\%$.
However, we suspect that theoretical modelling is also contributing to the difference w.r.t. our result.
While the newer data is not publicly available at this moment, we intend to extend the current analysis in a future communication.
In \cref{sec:app} we present the effects of simplifying the density profile from GBH to TH solution and fixing the model parameters such as $\Omega_m^{\rm bg}$ and $z_{\rm size}$, instead of marginalising over them to obtain the constraints on void contrast.

We chose to report the results of SN Ia analysis with and without the Planck CMB constraint on background matter density.
Both SN Ia datasets provide excellent fits in the case of a local underdensity, but do not exclude the possibility for a $z_{\rm size}>0.1$ overdensity (see \Cref{fig:GBHJLA,fig:GBHPanKBC}).
The Planck prior on background cosmology reduces the degeneracy between $\Omega_m^{\rm bg}$ and $\delta\Omega_0/\Omega_0$ parameters in SN Ia likelihood, providing tighter constraints.
Nevertheless, the main inferences presented here are not strongly affected by this prior due to the excellent agreement between the Planck value and our SN Ia best-fits for $\Omega_m^{\rm bg}$.
Examples presented in \cref{sec:app} also explore the case of using a fixed value for the relative matter density at the background, as contrary to sampling this likelihood parameter.

Compared to homogeneous model, the additional three parameters that characterise the void profile afford better fits to the SN Ia data, namely $\chi^2_{\rm JLA}=678.30$ and $\chi^2_{\rm Pan}=1020.72$ for $\Lambda$LTB model in contrast to $\chi^2_{\rm JLA}=682.94$ and $\chi^2_{\rm Pan}=1026.05$ for $\Lambda$CDM model.
We quantify this comparison with Akaike information criteria \cite{Akaike74} that penalises the usage of three extra parameters as $\Delta{\rm AIC}=\Delta\chi^2 + 2*3$.
All fits reported in \Cref{tab:1} have $\Delta{\rm AIC}<2$ implying no preference between the two models.

\subsection{Anisotropy considerations}
Using galaxy redshift distribution and number counts up to the depth of $z<0.1$, WS14 observed $215{h_{70}}^{-1}$Mpc local underdensity and an angular variation of its estimated density contrast in the range $-4\%$ to $-40\%$, with the mean of $-15\%$ (corresponds to $\delta\Omega_0/\Omega_0=-19\%$ shown in \Cref{fig:GBHJLA,fig:GBHPanKBC}).
Likewise, the Fig. 11a in KBC13 clearly shows discrepancy between LD points observed in different regions on the sky, again pointing towards a possibly anisotropic local density profile.
Hence, an important extension to the analysis presented here can be done by taking into account the anisotropy of the data, as well as the matter density profile and the observer's position inside of the considered inhomogeneity (see e.g. \cite{Blomqvist10}).
If there was any anisotropy present in the matter distribution, it would be averaged out in the analysis that uses isotropic model like ours, leaving us with the expectation to find angular variation of density contrast around this average when modelling for anisotropic matter profiles.
Be that as it may, the statistical significance in favour of the observed anisotropy in LD data is limited by the sky coverage of the used surveys, which amounts to $\sim$\,$15\%$ for KBC13 and $\sim$\,$22\%$ for WS14, spread over three distant patches on the celestial plane.

Several groups performed tests on SN Ia data looking for features of local and/or global anisotropy \citep{Huterer17,Bengaly15,Kalus13,Wang14,Sun18,Andrade18}, also in relation to the local matter distribution \citep{Colin11}.
These anisotropy estimates are also impaired by the inhomogeneous angular distribution of the observed SN on the sky, which is particularly visible on medium and high redshifts due to the narrow fields of the surveys, such as SDSS.
Probing low-redshift SN Ia for signatures of anisotropy, we estimate the statistical variance of model parameters over angular directions.
Concretely, the two parameters of interest, $\cal M$ and $\delta\Omega_0/\Omega_0$ in $\Lambda$LTB model, are tested separately for dipole anisotropy (see also e.g. \cite{Mariano12}).
The magnitude dipole is modelled as a vector of intensity ${\cal M}_d$ and direction $\bf n_d$.
Then, the magnitude parameter ${\cal M}^{(i)}$ of each SN Ia inside the void becomes  
\be
\label{eq:Md}
\begin{split}
{\cal M}^{(i)}&=\bar{\cal M}+{\cal M}_d\cos({\bf n^{(i)}, n_d})&&\text{for }z^{(i)}\leq z_{\rm size}\,,\\
{\cal M}^{(i)}&=\bar{\cal M}&&\text{for }z^{(i)}> z_{\rm size}\,.
\end{split}
\ee
where $\bar{\cal M}$ is the average intercept magnitude, while the normalised vector $\bf n^{(i)}$ represents the angular position of each SN Ia.
Although we fitted the full Pantheon dataset, the dipole is constrained mainly by 190 SN Ia under $z\leq0.08$, which are also more homogeneously distributed on the celestial plane compared to the full dataset.
Here we used a TH matter density profile for simplicity (see \cref{sec:th}).
Finally our estimate for the magnitude dipole
\be
{\cal M}_d=0.011^{+0.026}_{-0.031}
\ee
would produce the angular variation
\be
\Delta{\cal M}_{\rm r.m.s.}=\sqrt{Var({\Delta\cal M})}={\sqrt2\over2}{\cal M}_d=0.008\pm0.018\,.
\ee
This angular variance of magnitude inside the void can be related to the anisotropy of local expansion rate as
\be
\left(\frac{\Delta H_0^{\rm loc}}{H_0^{\rm loc}}\right)_{\rm r.m.s.}\approx0.2\ln(10)\,\,\Delta{\cal M}_{\rm r.m.s.}=0.4\pm0.8\%
\ee
Our result for the level of local anisotropy in Pantheon SN Ia is consistent with previous works that find null evidence against assumption of the isotropy \citep{Andrade18,ZhaoD19}, although less stringent compared to global anisotropy constraint found by \cite{Soltis19}.
Analogously to \cref{eq:Md} we statistically probed the dipole anisotropy of the estimated void contrast and found that it could account for the angular variance
\be
\Delta\left(\frac{\delta\Omega_0}{\Omega_0}\right)_{\rm r.m.s.}=3.7\pm4.1\%\,.
\ee
While this result at face value shows that Pantheon dataset does not exclude the angular variation of estimated void contrast $\Delta(\delta\Omega_0/\Omega_0)\sim$\,$10\%$ at $2\sigma$ confidence level, we still find it difficult to coincide our constraints on matter density profile from Pantheon SN Ia and LD data, or explain the level of anisotropy observed by WS14.
We would like to note that two former examples should be considered only as rough estimates of anisotropy level since: the dipole model does not represent a complete relativistic solution of FE, the variation of profile size is not considered, and the Pantheon SN Ia have had the apparent magnitudes corrected using the BBC method which relies on the assumption of isotropy \citep{Scolnic17}.
Although crucial in future studies, the present data is not quantitatively sufficing to benefit from the full anisotropic analysis given as an extension of the cosmological metric and theoretical model for the anisotropic density profile.

\subsection{Distance ladder measurements}
One of the most important implications of a large local void is the effect on direct measurement of present cosmic expansion rate by the means of low-redshift SN Ia.
Consider a spherical matter inhomogeneity, characterised with a density contrast of e.g. $\delta\Omega_0/\Omega_0=-25\%$ which is inside the $1\sigma$ confidence region of Pan+P constraints, presented in \Cref{tab:1} and \Cref{fig:GBHPanKBC}.
This matter contrast will also induce a higher local present expansion rate by $\delta H_0/H_0\approx3.6\%$ w.r.t. the background expansion rate (see \Cref{eq:dOm,eq:dH}.
Hence, the luminosity distance of a SN located on any shell inside the local void is higher than what one would estimate using the background metric.
This can affect the analysis of SN Ia data in the redshift range inside the local void as well as the resulting model inferences.

In light of everything said above, let's assume our local matter distribution has indeed an isotropic underdense profile.
Analysing SN Ia data from the redshift range inside this local void with a simpler homogeneous model would provide a different constraint on $H_0^{\rm bg}$ than the LTB model.
In fact, the assumption of locally homogeneous cosmic expansion ($H_0^{\rm loc}=H_0^{\rm bg}$) can overestimate the value of background expansion rate, while underestimating $H_0^{\rm loc}$, by $\sim\frac12 \delta H_0/H_0$ and affect the inferences on other background cosmological parameters.

As discussed in \Cref{ssec:resLD} theoretical expectations for the large local void based on $\Lambda$CDM model, and its effect on $H_0$ measurement are small, which is reflected in results presented by recent studies.
On one side, the level of expected cosmic variance in local matter density for the Planck $\Lambda$CDM model is estimated to impact the systematic error of $H_0^{\rm bg}$ at $\gtrsim$\,$2\%$ when using SN Ia from $z\geq0.01$ and at $\gtrsim$\,$1\%$ when using SN Ia from $z\geq0.0233$ \citep{Marra13}.
On the other side, the large-volume cosmological N-body simulations based on the Planck $\Lambda$CDM model, such as e.g. in \cite{Skillman14}, were used to quantify effects of both the cosmic variance and the SN Ia sample selection on $H_0^{\rm bg}$ value - \cite{Wu17,Odderskov16} estimate the expected variance level of measured $H_0^{\rm bg}$ to be $\sim$\,$0.5\%$ for $0.0233<z<0.15$ cut (c.f. an earlier work by \cite{Wojtak14}).
In order to minimise the effect of local matter distribution, R16 (including their previous works) use {\it only} the SN Ia above $z>0.0233$ for the model-independent estimate of $H_0^{\rm bg}$ \citep{Jha07}.
In consistence with \cite{Wu17}, KSR19 report no evidence for strong variation of matter density profile in the range $z<0.15$ from SN Ia data, securing the ability to measure $H_0^{\rm bg}$ value to a percent value.

The measurement of background expansion rate based on low-redshift SN Ia is obtained from the intercept magnitude parameter $\cal M$ (see \Cref{ssec:resSN}).
While the exact value of $H_0^{\rm bg}$ relies on local SN Ia luminosity calibration, we are mainly interested at the relative change of this estimate due to the effect of a large local void, which can be seen as the shift in estimated value of $\cal M$.
For this purpose, and only in this subsection, we analyse 237 Pantheon SN Ia from $0.0233< z<0.15$ redshift cut.
Fitting them to FLRW series expansion (SE) formula for luminosity distance and using a fixed value of $q_0=-0.55$, as presented in R16, we obtain ${\cal M}({\rm SE})=-1.1825\pm0.0094$.
In low redshift range, this fit is equivalent to $\Lambda$CDM fit with $\Omega_m=0.3$, which we will use for comparison to underdense scenario.
The same Pantheon cut is fitted to $\Lambda$LTB model with background cosmology prior $\Omega_m^{\rm bg}=0.3$ ($q_0^{\rm bg}=-0.55$), density contrast prior range $-50\%\leq\delta\Omega_0/\Omega_0\leq5\%$, void size $0.025\leq z_{\rm size}\leq0.12$, and shape parameter $0\leq\Delta z\leq z_{\rm size}$.
In order to ensure enough SN Ia that constrain the background, we consider only profiles that reach the background regime up to $z\leq0.13$ by additionally removing the region of parameter space with such combination of large $z_{\rm size}$ and $\Delta z$ values.
After marginalising the matter density profile out of the low-redshift Pantheon likelihood, we estimate the intercept SN magnitude parameter\footnote{Confidence intervals are given between 16th and 84th percentiles} to be
${\cal M}({\rm \Lambda LTB})=-1.1577_{-0.0148}^{+0.0158}$.
The remaining parameter constraints are presented in the last row of \Cref{tab:1}.
Regardless the three additional parameters that model the local matter underdesnity, the $\Lambda$LTB fit to low-redshift Pantheon SN Ia reveals an improvement of the best-fit by $\Delta$AIC=-3.3.

\begin{figure}
    \includegraphics[width=.48\textwidth]{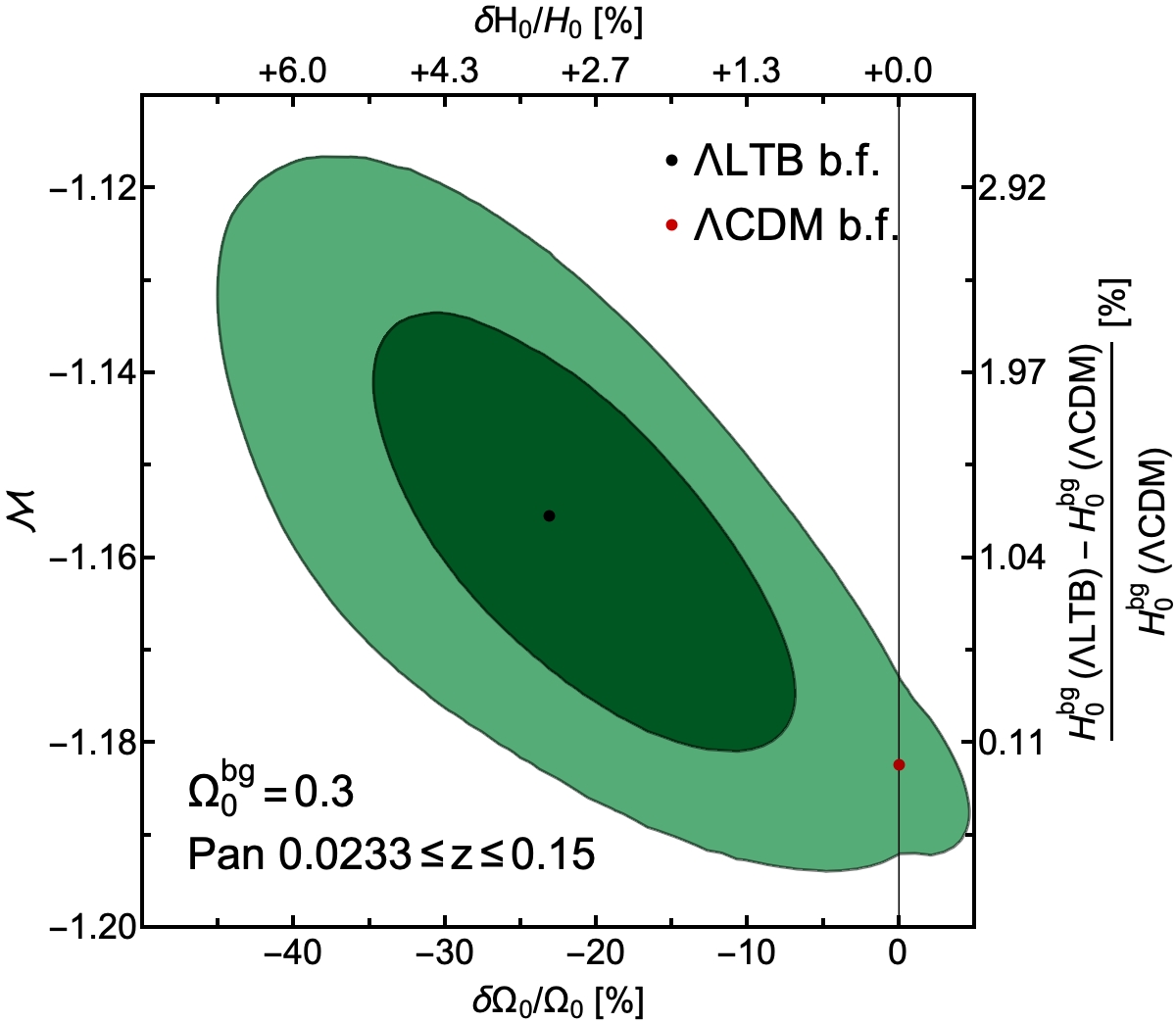}
    \caption{$1\sigma$ and $2\sigma$ confidence regions for $\Lambda$LTB model with GBH density profile, fitted to 237 low-redshift ($0.0233<z<0.15$) SN Ia from Pantheon dataset. The best fit (b.f.) for $\Lambda$CDM model, shown as red point is almost $2\sigma$ away from $\Lambda$LTB constraints.}
    \label{fig:GBHPanr}
\end{figure}

The local matter underdensity affects the estimated value of $\cal M$ due to its degenearcy with density constrast.
This can be seen on \Cref{fig:GBHPanr}, where we show the low-redshift Pantheon constraints in $\delta\Omega_0/\Omega_0-{\cal M}$ plane.
The best-fit of $\Lambda$CDM model (equivalent to SE) can be found on vertical axis for zero void contrast and is marked as a red point in \Cref{fig:GBHPanr}, which is $1.7\sigma$ away from our constraints on $\Lambda$LTB model.
Both values for the intercept magnitude parameter $\cal M$ have been estimated fitting the same 237 Pantheon SN Ia to two different models and are, hence, fully correlated.
Taking into account this correlation, the difference between two values is evalueted to be $\Delta{\cal M}=0.0247^{+0.0065}_{-0.0055}$.
As the $\Lambda$LTB model has more parameters in fitting, the error bar of resulting ${\cal M}({\rm \Lambda LTB})$ is bigger than in the case of ${\cal M}({\rm \Lambda CDM})$, and elseways, we note that if they were equivalent then the uncertainty on estimated shift $\Delta{\cal M}$ would have been zero.
Since this change of intercept magnitude value is coming only from the theoretical modelling and the two models have comperable AIC values (see \Cref{tab:1}), it should be seen as a possible source of systematic error on $\cal M$ estimate.
The positive shift of $\Delta{\cal M}=0.0247$ corresponds to $1.14\%$ lower $H_0^{\rm bg}$ obtained with $\Lambda$LTB model than with SE formula.
Compared to the total error of ${\cal M}({\rm SE})$, this value of $\Delta{\cal M}$ is significant.
However, the major contribution to the uncertainty of $H_0$ measurement is presently constricted to calibration of standard SN Ia absolute magnitude, $M_B$, that together with much smaller uncertainty on intercept magnitude $\cal M(\rm SE)$ reaches the total error bar of $2.4\%$ reported by R16.
Adding the effect of large local matter inhomogeneity ($z_{\rm size}>0.025$) on estimate of SN Ia intercept magnitude could increase this uncertainty level to $2.7\%$ on the lower end of $H_0$ measurement.
The improvement in Cepheid and SN Ia calibation techniques {\bf(see R19)}, followed by reduction of uncertainty on $M_B$, can lead to a situation in which the effect of local matter inhomogeneity becomes more relevant for the measurement of background cosmic expansion rate.

\section{Conclusions}
\label{sec:con}
In the present work we have investigated the evidence for a large local void using a well formulated $\Lambda$LTB model against the most recent publicly available SN Ia compilations.
In order to reduce the correlation with background cosmology, we have used the CMB constraint on $\Omega_m^{\rm bg}$ \citep{Planck18}, which aids to provide a slightly tighter constraints on the local void profile.
The analysis on SN Ia datasets is also complimented with the luminosity density data obtained from galaxy surveys by \cite{Keenan13}, which together with \cite{Whitbourn14} played a key role in propelling the recent discussion about possible existence of a large local void and its effect towards resolving one of the most prominent discordances in modern cosmology - the $H_0$ tension.
We summarise our primary results in the following.
All reported confidence intervals (c.i.) are between 16th and 84th percentiles.
\begin{itemize}
\item Fitting $\Lambda$LTB model to LD data from KBC13, we find a void with density contrast of $\left(\delta\rho_0/\rho_0\right)^{c.i.} = -43.8\pm 6.0\%$, deeper than originally proposed $\delta\rho_0/\rho_0 = -30\%$.
\item From SN Ia we do not find a strong evidence for a void or otherwise -- a fit to Pan+P yields $\left(\delta\rho_0/\rho_0\right)^{c.i.} = -7.9_{-5.8}^{+7.0}\%$ which corresponds to $\left(\delta\Omega_0/\Omega_0\right)^{c.i.}=-10.5^{+9.3}_{-7.4}\%$, and a wide range for the redshift size $z_{\rm size}^{c.i.}=0.068^{+0.021}_{-0.030}$.
However, Pantheon constraints do not exclude a void of e.g. $z_{\rm size}\approx0.075$ and $\delta\Omega_0/\Omega_0\approx-25\%$ at $1\sigma$ confidence level (c.f. \Cref{fig:GBHPanKBC}).
\item JLA likelihood is in a good overall agreement with Pantheon, except in the range $z\leq0.04$ (c.f. \Cref{fig:GBHJLA,fig:GBHPanKBC}). Both SN Ia datasets are at $\gtrsim$\,$3\sigma$ tension with our result obtained from KBC13 data, but in excellent agreement with parameters of the isotropic void proposed by WS14.
\item Leave one out analysis of KBC13 data reveals that the corresponding estimated contrast is dominantly constrained by one stringent point.
Removing this only data point from 2MASS survey and using the remaining 9 points from UKIDSS and GAMA surveys, relaxes their tension with SN Ia data.
\item The model comparison with Akaike information criteria shows no preference between the tested models (c.f. \Cref{tab:1}).
\item Probing the level of statistical variance in $\Lambda$LTB model parameters over angular direction, we find null evidence for the dipole anisotropy.
Its contribution to the angular variance is small: $\Delta{\cal M}_{\rm r.m.s.}=0.008\pm0.018$ for the SN intercept magnitude parameter or $\Delta\left(\frac{\delta\Omega_0}{\Omega_0}\right)_{\rm r.m.s.}=3.7\pm4.1\%$ for the void contrast.
\item In our analysis of SN Ia data we do not find any evidence for a large isotropic void that could resolve the $9\%$ discrepancy between two $H_0$ estimates in tension, leaving this local effect alone to be a highly unlikely explanation.
\item We fitted low-redshift range of Pantheon SN Ia data from $0.0233<z<0.15$ to the FLRW series expansion formula for luminosity distance and to the $\Lambda$LTB model, comparing their estimates of SN Ia intercept magnitude, which is used for $H_0$ measurement.
Adding the possibility of a large local void, we observe a small improvement in the fit by $\Delta AIC=-3.3$ and a shift of the inferred intercept magnitude parameter $\Delta{\cal M}=0.0247^{+0.065}_{-0.055}$ that could be a source of additional systematic error in $H_0^{\rm bg}$ of $1.14\%$.
While the simpler model, which is based on series expansion formula of luminosity distance in FLRW metric, is only $1.7\sigma$ away from our $\Lambda$LTB constraints (see \cref{fig:GBHPanr}), this result is questioning whether the total error budget of distance ladder measurements based on low-redshift SN Ia should be reconsidered.
\end{itemize}

Increase of presently available data, followed by more complex theoretical modelling, is necessary to better understand the observed disagreement between LD and SN Ia samples.
As already mentioned, considering off-centre position of the observer would extend the present isotropic $\Lambda$LTB formalism, also allowing for an anisotropic point of view by construction.
With increasing number of observations in future surveys, SN Ia may prove as effective tracers of local matter density distribution.

\section*{Acknowledgements}
The authors would like to thank Amy J. Barger, Benjamin L. Hoscheit, Adam Riess and Tom Shanks for useful discussions, and Caterina Traficante for contribution to this work during her thesis. Authors acknowledge financial support by ASI Grant No. 2016-24-H.0.

\bibliography{references}

\begin{thebibliography}{}
\makeatletter
\relax
\def\mn@urlcharsother{\let\do\@makeother \do\$\do\&\do\#\do\^\do\_\do\%\do\~}
\def\mn@doi{\begingroup\mn@urlcharsother \@ifnextchar [ {\mn@doi@}
  {\mn@doi@[]}}
\def\mn@doi@[#1]#2{\def\@tempa{#1}\ifx\@tempa\@empty \href
  {http://dx.doi.org/#2} {doi:#2}\else \href {http://dx.doi.org/#2} {#1}\fi
  \endgroup}
\def\mn@eprint#1#2{\mn@eprint@#1:#2::\@nil}
\def\mn@eprint@arXiv#1{\href {http://arxiv.org/abs/#1} {{\tt arXiv:#1}}}
\def\mn@eprint@dblp#1{\href {http://dblp.uni-trier.de/rec/bibtex/#1.xml}
  {dblp:#1}}
\def\mn@eprint@#1:#2:#3:#4\@nil{\def\@tempa {#1}\def\@tempb {#2}\def\@tempc
  {#3}\ifx \@tempc \@empty \let \@tempc \@tempb \let \@tempb \@tempa \fi \ifx
  \@tempb \@empty \def\@tempb {arXiv}\fi \@ifundefined
  {mn@eprint@\@tempb}{\@tempb:\@tempc}{\expandafter \expandafter \csname
  mn@eprint@\@tempb\endcsname \expandafter{\@tempc}}}

\bibitem[\protect\citeauthoryear{{Akaike}}{{Akaike}}{1974}]{Akaike74}
{Akaike} H.,  1974, IEEE Transactions on Automatic Control, \href
  {http://adsabs.harvard.edu/abs/1974ITAC...19..716A} {19, 716}

\bibitem[\protect\citeauthoryear{{Alam} et~al.,}{{Alam} et~al.}{2017}]{Alam17}
{Alam} S.,  et~al., 2017, \mn@doi [\mnras] {10.1093/mnras/stx721}, \href
  {http://adsabs.harvard.edu/abs/2017MNRAS.470.2617A} {470, 2617}

\bibitem[\protect\citeauthoryear{{Alfedeel} \& {Hellaby}}{{Alfedeel} \&
  {Hellaby}}{2010}]{Alfedeel10}
{Alfedeel} A. H.~A.,  {Hellaby} C.,  2010, \mn@doi [General Relativity and
  Gravitation] {10.1007/s10714-010-0971-y}, \href
  {https://ui.adsabs.harvard.edu/abs/2010GReGr..42.1935A} {42, 1935}

\bibitem[\protect\citeauthoryear{{Alnes}, {Amarzguioui}  \& {Gr{\o}n}}{{Alnes}
  et~al.}{2006}]{Alnes06a}
{Alnes} H.,  {Amarzguioui} M.,   {Gr{\o}n} {\O}.,  2006, \mn@doi [\prd]
  {10.1103/PhysRevD.73.083519}, \href
  {http://adsabs.harvard.edu/abs/2006PhRvD..73h3519A} {73, 083519}

\bibitem[\protect\citeauthoryear{{Amendola}, {Eggers Bj{\ae} lde}, {Valkenburg}
   \& {Wong}}{{Amendola} et~al.}{2013}]{Amendola13}
{Amendola} L.,  {Eggers Bj{\ae} lde} O.,  {Valkenburg} W.,   {Wong} Y.~Y.~Y.,
  2013, \mn@doi [\jcap] {10.1088/1475-7516/2013/12/042}, \href
  {http://adsabs.harvard.edu/abs/2013JCAP...12..042A} {12, 042}

\bibitem[\protect\citeauthoryear{{Andrade}, {Bengaly}, {Santos}  \&
  {Alcaniz}}{{Andrade} et~al.}{2018}]{Andrade18}
{Andrade} U.,  {Bengaly} C. A.~P.,  {Santos} B.,   {Alcaniz} J.~S.,  2018,
  \mn@doi [\apj] {10.3847/1538-4357/aadb90}, \href
  {https://ui.adsabs.harvard.edu/abs/2018ApJ...865..119A} {865, 119}

\bibitem[\protect\citeauthoryear{{Aubourg} et~al.,}{{Aubourg}
  et~al.}{2015}]{Aubourg15}
{Aubourg} {\'E}.,  et~al., 2015, \mn@doi [\prd] {10.1103/PhysRevD.92.123516},
  \href {https://ui.adsabs.harvard.edu/abs/2015PhRvD..92l3516A} {92, 123516}

\bibitem[\protect\citeauthoryear{{Bautista} et~al.,}{{Bautista}
  et~al.}{2017}]{Bautista17}
{Bautista} J.~E.,  et~al., 2017, \mn@doi [\aap] {10.1051/0004-6361/201730533},
  \href {http://adsabs.harvard.edu/abs/2017A%26A...603A..12B} {603, A12}

\bibitem[\protect\citeauthoryear{{Beaton} et~al.,}{{Beaton}
  et~al.}{2016}]{Beaton16}
{Beaton} R.~L.,  et~al., 2016, \mn@doi [\apj] {10.3847/0004-637X/832/2/210},
  \href {http://adsabs.harvard.edu/abs/2016ApJ...832..210B} {832, 210}

\bibitem[\protect\citeauthoryear{{Bengaly}, {Bernui}  \& {Alcaniz}}{{Bengaly}
  et~al.}{2015}]{Bengaly15}
{Bengaly} C.~A.~P. J.,  {Bernui} A.,   {Alcaniz} J.~S.,  2015, \mn@doi [\apj]
  {10.1088/0004-637X/808/1/39}, \href
  {https://ui.adsabs.harvard.edu/abs/2015ApJ...808...39B} {808, 39}

\bibitem[\protect\citeauthoryear{{Betoule} et~al.,}{{Betoule}
  et~al.}{2014}]{Betoule14}
{Betoule} M.,  et~al., 2014, \mn@doi [\aap] {10.1051/0004-6361/201423413},
  \href {http://adsabs.harvard.edu/abs/2014A%26A...568A..22B} {568, A22}

\bibitem[\protect\citeauthoryear{{Birrer} et~al.,}{{Birrer}
  et~al.}{2019}]{Birrer19}
{Birrer} S.,  et~al., 2019, \mn@doi [\mnras] {10.1093/mnras/stz200}, \href
  {https://ui.adsabs.harvard.edu/abs/2019MNRAS.484.4726B} {484, 4726}

\bibitem[\protect\citeauthoryear{{Blomqvist} \& {M{\"o}rtsell}}{{Blomqvist} \&
  {M{\"o}rtsell}}{2010}]{Blomqvist10}
{Blomqvist} M.,  {M{\"o}rtsell} E.,  2010, \mn@doi [\jcap]
  {10.1088/1475-7516/2010/05/006}, \href
  {http://adsabs.harvard.edu/abs/2010JCAP...05..006B} {5, 006}

\bibitem[\protect\citeauthoryear{{Boehringer}, {Chon}  \&
  {Collins}}{{Boehringer} et~al.}{2019}]{Boehringer19}
{Boehringer} H.,  {Chon} G.,   {Collins} C.~A.,  2019, arXiv e-prints, \href
  {https://ui.adsabs.harvard.edu/abs/2019arXiv190712402B} {p. arXiv:1907.12402}

\bibitem[\protect\citeauthoryear{Bolejko \& Sussman}{Bolejko \&
  Sussman}{2011}]{Bolejko11}
Bolejko K.,  Sussman R.~A.,  2011, \mn@doi [Physics Letters B]
  {10.1016/j.physletb.2011.02.007}, 697, 265

\bibitem[\protect\citeauthoryear{{Bondi}}{{Bondi}}{1947}]{Bondi47}
{Bondi} H.,  1947, \mn@doi [\mnras] {10.1093/mnras/107.5-6.410}, \href
  {http://adsabs.harvard.edu/abs/1947MNRAS.107..410B} {107, 410}

\bibitem[\protect\citeauthoryear{{Camarena} \& {Marra}}{{Camarena} \&
  {Marra}}{2018}]{Camarena18}
{Camarena} D.,  {Marra} V.,  2018, \mn@doi [\prd] {10.1103/PhysRevD.98.023537},
  \href {https://ui.adsabs.harvard.edu/abs/2018PhRvD..98b3537C} {98, 023537}

\bibitem[\protect\citeauthoryear{{C{\'e}l{\'e}rier}}{{C{\'e}l{\'e}rier}}{2000}]{Celerier00}
{C{\'e}l{\'e}rier} M.-N.,  2000, \aap, \href
  {http://adsabs.harvard.edu/abs/2000A%26A...353...63C} {353, 63}

\bibitem[\protect\citeauthoryear{{Chen}, {Kumar}  \& {Ratra}}{{Chen}
  et~al.}{2017}]{ChenRatra17}
{Chen} Y.,  {Kumar} S.,   {Ratra} B.,  2017, \mn@doi [\apj]
  {10.3847/1538-4357/835/1/86}, \href
  {https://ui.adsabs.harvard.edu/abs/2017ApJ...835...86C} {835, 86}

\bibitem[\protect\citeauthoryear{{Clifton}, {Ferreira}  \& {Land}}{{Clifton}
  et~al.}{2008}]{Clifton08}
{Clifton} T.,  {Ferreira} P.~G.,   {Land} K.,  2008, \mn@doi [\prl]
  {10.1103/PhysRevLett.101.131302}, \href
  {https://ui.adsabs.harvard.edu/abs/2008PhRvL.101m1302C} {101, 131302}

\bibitem[\protect\citeauthoryear{{Colin}, {Mohayaee}, {Sarkar}  \&
  {Shafieloo}}{{Colin} et~al.}{2011}]{Colin11}
{Colin} J.,  {Mohayaee} R.,  {Sarkar} S.,   {Shafieloo} A.,  2011, \mn@doi
  [\mnras] {10.1111/j.1365-2966.2011.18402.x}, \href
  {https://ui.adsabs.harvard.edu/abs/2011MNRAS.414..264C} {414, 264}

\bibitem[\protect\citeauthoryear{{Cuceu}, {Farr}, {Lemos}  \&
  {Font-Ribera}}{{Cuceu} et~al.}{2019}]{Cuceu19}
{Cuceu} A.,  {Farr} J.,  {Lemos} P.,   {Font-Ribera} A.,  2019,
  arXiv:1906.11628, \href
  {https://ui.adsabs.harvard.edu/abs/2019arXiv190611628C} {}

\bibitem[\protect\citeauthoryear{{Driver} \& {Robotham}}{{Driver} \&
  {Robotham}}{2010}]{Driver10}
{Driver} S.~P.,  {Robotham} A. S.~G.,  2010, \mn@doi [\mnras]
  {10.1111/j.1365-2966.2010.17028.x}, \href
  {https://ui.adsabs.harvard.edu/abs/2010MNRAS.407.2131D} {407, 2131}

\bibitem[\protect\citeauthoryear{{Driver} et~al.,}{{Driver}
  et~al.}{2011}]{Driver11}
{Driver} S.~P.,  et~al., 2011, \mn@doi [\mnras]
  {10.1111/j.1365-2966.2010.18188.x}, \href
  {https://ui.adsabs.harvard.edu/abs/2011MNRAS.413..971D} {413, 971}

\bibitem[\protect\citeauthoryear{{Eisenstein} et~al.,}{{Eisenstein}
  et~al.}{2005}]{Eisenstein05}
{Eisenstein} D.~J.,  et~al., 2005, \mn@doi [\apj] {10.1086/466512}, \href
  {http://adsabs.harvard.edu/abs/2005ApJ...633..560E} {633, 560}

\bibitem[\protect\citeauthoryear{{February}, {Larena}, {Smith}  \&
  {Clarkson}}{{February} et~al.}{2010}]{February10}
{February} S.,  {Larena} J.,  {Smith} M.,   {Clarkson} C.,  2010, \mn@doi
  [\mnras] {10.1111/j.1365-2966.2010.16627.x}, \href
  {http://adsabs.harvard.edu/abs/2010MNRAS.405.2231F} {405, 2231}

\bibitem[\protect\citeauthoryear{{Fern{\'a}ndez Arenas} et~al.,}{{Fern{\'a}ndez
  Arenas} et~al.}{2018}]{Arenas18}
{Fern{\'a}ndez Arenas} D.,  et~al., 2018, \mn@doi [\mnras]
  {10.1093/mnras/stx2710}, \href
  {http://adsabs.harvard.edu/abs/2018MNRAS.474.1250F} {474, 1250}

\bibitem[\protect\citeauthoryear{{Freedman} \& {Madore}}{{Freedman} \&
  {Madore}}{2010}]{Freedman10}
{Freedman} W.~L.,  {Madore} B.~F.,  2010, \mn@doi [\araa]
  {10.1146/annurev-astro-082708-101829}, \href
  {http://adsabs.harvard.edu/abs/2010ARA%26A..48..673F} {48, 673}

\bibitem[\protect\citeauthoryear{{Freedman} et~al.,}{{Freedman}
  et~al.}{2001}]{Freedman01}
{Freedman} W.~L.,  et~al., 2001, \mn@doi [\apj] {10.1086/320638}, \href
  {http://adsabs.harvard.edu/abs/2001ApJ...553...47F} {553, 47}

\bibitem[\protect\citeauthoryear{{Freedman} et~al.,}{{Freedman}
  et~al.}{2019}]{Freedman19}
{Freedman} W.~L.,  et~al., 2019, arXiv:1907.05922, \href
  {https://ui.adsabs.harvard.edu/abs/2019arXiv190705922F} {}

\bibitem[\protect\citeauthoryear{{Garcia-Bellido} \&
  {Haugb{\o}lle}}{{Garcia-Bellido} \& {Haugb{\o}lle}}{2008}]{Garcia08}
{Garcia-Bellido} J.,  {Haugb{\o}lle} T.,  2008, \mn@doi [J. Cosmol. Astropart.
  Phys.] {10.1088/1475-7516/2008/04/003}, \href
  {https://ui.adsabs.harvard.edu/abs/2008JCAP...04..003G} {2008, 003}

\bibitem[\protect\citeauthoryear{{G{\'o}mez-Valent}}{{G{\'o}mez-Valent}}{2019}]{Gomez-Valent18}
{G{\'o}mez-Valent} A.,  2019, \mn@doi [\jcap] {10.1088/1475-7516/2019/05/026},
  \href {https://ui.adsabs.harvard.edu/abs/2019JCAP...05..026G} {2019, 026}

\bibitem[\protect\citeauthoryear{{Haridasu}, {Lukovi{\'c}}, {D'Agostino}  \&
  {Vittorio}}{{Haridasu} et~al.}{2017}]{Haridasu17}
{Haridasu} B.~S.,  {Lukovi{\'c}} V.~V.,  {D'Agostino} R.,   {Vittorio} N.,
  2017, \mn@doi [\aap] {10.1051/0004-6361/201730469}, \href
  {http://adsabs.harvard.edu/abs/2017A%26A...600L...1H} {600, L1}

\bibitem[\protect\citeauthoryear{Haridasu, Lukovi\'c  \& Vittorio}{Haridasu
  et~al.}{2018a}]{Haridasu17a}
Haridasu B.~S.,  Lukovi\'c V.~V.,   Vittorio N.,  2018a, \mn@doi [JCAP]
  {10.1088/1475-7516/2018/05/033}, 1805, 033

\bibitem[\protect\citeauthoryear{Haridasu, Lukovi\'c, Moresco  \&
  Vittorio}{Haridasu et~al.}{2018b}]{Haridasu18_GP}
Haridasu B.~S.,  Lukovi\'c V.~V.,  Moresco M.,   Vittorio N.,  2018b, \mn@doi
  [JCAP] {10.1088/1475-7516/2018/10/015}, 1810, 015

\bibitem[\protect\citeauthoryear{{Hinshaw} et~al.,}{{Hinshaw}
  et~al.}{2013}]{WMAP13}
{Hinshaw} G.,  et~al., 2013, \mn@doi [\apjs] {10.1088/0067-0049/208/2/19},
  \href {http://adsabs.harvard.edu/abs/2013ApJS..208...19H} {208, 19}

\bibitem[\protect\citeauthoryear{{Holsclaw}, {Alam}, {Sans{\'o}}, {Lee},
  {Heitmann}, {Habib}  \& {Higdon}}{{Holsclaw} et~al.}{2010}]{Holsclaw10}
{Holsclaw} T.,  {Alam} U.,  {Sans{\'o}} B.,  {Lee} H.,  {Heitmann} K.,  {Habib}
  S.,   {Higdon} D.,  2010, \mn@doi [\prl] {10.1103/PhysRevLett.105.241302},
  \href {https://ui.adsabs.harvard.edu/abs/2010PhRvL.105x1302H} {105, 241302}

\bibitem[\protect\citeauthoryear{{Hoscheit} \& {Barger}}{{Hoscheit} \&
  {Barger}}{2017}]{Hoscheit17}
{Hoscheit} B.~L.,  {Barger} A.~J.,  2017, in American Astronomical Society
  Meeting Abstracts \#230. p. 314.05

\bibitem[\protect\citeauthoryear{{Huterer}, {Shafer}, {Scolnic}  \&
  {Schmidt}}{{Huterer} et~al.}{2017}]{Huterer17}
{Huterer} D.,  {Shafer} D.~L.,  {Scolnic} D.~M.,   {Schmidt} F.,  2017, \mn@doi
  [J. Cosmol. Astropart. Phys.] {10.1088/1475-7516/2017/05/015}, \href
  {https://ui.adsabs.harvard.edu/abs/2017JCAP...05..015H} {2017, 015}

\bibitem[\protect\citeauthoryear{{Jha}, {Riess}  \& {Kirshner}}{{Jha}
  et~al.}{2007}]{Jha07}
{Jha} S.,  {Riess} A.~G.,   {Kirshner} R.~P.,  2007, \mn@doi [\apj]
  {10.1086/512054}, \href
  {https://ui.adsabs.harvard.edu/abs/2007ApJ...659..122J} {659, 122}

\bibitem[\protect\citeauthoryear{{Jimenez} \& {Loeb}}{{Jimenez} \&
  {Loeb}}{2002}]{Jimenez02}
{Jimenez} R.,  {Loeb} A.,  2002, \mn@doi [\apj] {10.1086/340549}, \href
  {http://adsabs.harvard.edu/abs/2002ApJ...573...37J} {573, 37}

\bibitem[\protect\citeauthoryear{{Jimenez}, {Cimatti}, {Verde}, {Moresco}  \&
  {Wandelt}}{{Jimenez} et~al.}{2019}]{Jimenez19}
{Jimenez} R.,  {Cimatti} A.,  {Verde} L.,  {Moresco} M.,   {Wandelt} B.,  2019,
  \mn@doi [\jcap] {10.1088/1475-7516/2019/03/043}, \href
  {https://ui.adsabs.harvard.edu/abs/2019JCAP...03..043J} {2019, 043}

\bibitem[\protect\citeauthoryear{{Joudaki} et~al.,}{{Joudaki}
  et~al.}{2019}]{Joudaki19}
{Joudaki} S.,  et~al., 2019, arXiv:1906.09262, \href
  {https://ui.adsabs.harvard.edu/abs/2019arXiv190609262J} {}

\bibitem[\protect\citeauthoryear{{Kalus}, {Schwarz}, {Seikel}  \&
  {Wiegand}}{{Kalus} et~al.}{2013}]{Kalus13}
{Kalus} B.,  {Schwarz} D.~J.,  {Seikel} M.,   {Wiegand} A.,  2013, \mn@doi
  [\aap] {10.1051/0004-6361/201220928}, \href
  {https://ui.adsabs.harvard.edu/abs/2013A&A...553A..56K} {553, A56}

\bibitem[\protect\citeauthoryear{{Keenan}, {Barger}, {Cowie}, {Wang}, {Wold}
  \& {Trouille}}{{Keenan} et~al.}{2012}]{Keenan12}
{Keenan} R.~C.,  {Barger} A.~J.,  {Cowie} L.~L.,  {Wang} W.~H.,  {Wold} I.,
  {Trouille} L.,  2012, \mn@doi [\apj] {10.1088/0004-637X/754/2/131}, \href
  {https://ui.adsabs.harvard.edu/abs/2012ApJ...754..131K} {754, 131}

\bibitem[\protect\citeauthoryear{{Keenan}, {Barger}  \& {Cowie}}{{Keenan}
  et~al.}{2013}]{Keenan13}
{Keenan} R.~C.,  {Barger} A.~J.,   {Cowie} L.~L.,  2013, \mn@doi [\apj]
  {10.1088/0004-637X/775/1/62}, \href
  {http://adsabs.harvard.edu/abs/2013ApJ...775...62K} {775, 62}

\bibitem[\protect\citeauthoryear{{Kenworthy}, {Scolnic}  \&
  {Riess}}{{Kenworthy} et~al.}{2019}]{Kenworthy19}
{Kenworthy} W.~D.,  {Scolnic} D.,   {Riess} A.,  2019, \mn@doi [\apj]
  {10.3847/1538-4357/ab0ebf}, \href
  {https://ui.adsabs.harvard.edu/abs/2019ApJ...875..145K} {875, 145}

\bibitem[\protect\citeauthoryear{Krasi\'{n}ski}{Krasi\'{n}ski}{1997}]{Krasinski97}
Krasi\'{n}ski A.,  1997, Inhomogeneous Cosmological Models.
Cambridge University Press, \mn@doi{10.1017/CBO9780511721694}

\bibitem[\protect\citeauthoryear{{Lange}, {Yang}, {Guo}, {Luo}  \& {van den
  Bosch}}{{Lange} et~al.}{2019}]{Lange19}
{Lange} J.~U.,  {Yang} X.,  {Guo} H.,  {Luo} W.,   {van den Bosch} F.~C.,
  2019, arXiv:1906.08680, \href
  {https://ui.adsabs.harvard.edu/abs/2019arXiv190608680L} {}

\bibitem[\protect\citeauthoryear{{Lavaux} \& {Hudson}}{{Lavaux} \&
  {Hudson}}{2011}]{Lavaux11}
{Lavaux} G.,  {Hudson} M.~J.,  2011, \mn@doi [\mnras]
  {10.1111/j.1365-2966.2011.19233.x}, \href
  {https://ui.adsabs.harvard.edu/abs/2011MNRAS.416.2840L} {416, 2840}

\bibitem[\protect\citeauthoryear{{Lawrence} et~al.,}{{Lawrence}
  et~al.}{2007}]{Lawrence07}
{Lawrence} A.,  et~al., 2007, \mn@doi [\mnras]
  {10.1111/j.1365-2966.2007.12040.x}, \href
  {https://ui.adsabs.harvard.edu/abs/2007MNRAS.379.1599L} {379, 1599}

\bibitem[\protect\citeauthoryear{{Lema{\^i}tre}}{{Lema{\^i}tre}}{1927}]{Lemaitre27}
{Lema{\^i}tre} G.,  1927, Annal. Soc. Sci. Brux., 47, 49

\bibitem[\protect\citeauthoryear{{Lema{\^i}tre}}{{Lema{\^i}tre}}{1933}]{Lemaitre33}
{Lema{\^i}tre} G.,  1933, Annales de la Soci{\'e}t{\'e} Scientifique de
  Bruxelles, \href {http://adsabs.harvard.edu/abs/1933ASSB...53...51L} {53, 51}

\bibitem[\protect\citeauthoryear{{Lemos}, {Lee}, {Efstathiou}  \&
  {Gratton}}{{Lemos} et~al.}{2019}]{Lemos19}
{Lemos} P.,  {Lee} E.,  {Efstathiou} G.,   {Gratton} S.,  2019, \mn@doi
  [\mnras] {10.1093/mnras/sty3082}, \href
  {https://ui.adsabs.harvard.edu/abs/2019MNRAS.483.4803L} {483, 4803}

\bibitem[\protect\citeauthoryear{{Liao}, {Fan}, {Ding}, {Biesiada}  \&
  {Zhu}}{{Liao} et~al.}{2017}]{Liao17}
{Liao} K.,  {Fan} X.-L.,  {Ding} X.,  {Biesiada} M.,   {Zhu} Z.-H.,  2017,
  \mn@doi [Nature Communications] {10.1038/s41467-017-01152-9}, \href
  {https://ui.adsabs.harvard.edu/abs/2017NatCo...8.1148L} {8, 1148}

\bibitem[\protect\citeauthoryear{{Lukovi{\'c}}, {D'Agostino}  \&
  {Vittorio}}{{Lukovi{\'c}} et~al.}{2016}]{Lukovic16}
{Lukovi{\'c}} V.~V.,  {D'Agostino} R.,   {Vittorio} N.,  2016, \mn@doi [\aap]
  {10.1051/0004-6361/201628217}, \href
  {http://adsabs.harvard.edu/abs/2016A%26A...595A.109L} {595, A109}

\bibitem[\protect\citeauthoryear{{Macaulay} et~al.,}{{Macaulay}
  et~al.}{2019}]{Macaulay19}
{Macaulay} E.,  et~al., 2019, \mn@doi [\mnras] {10.1093/mnras/stz978}, \href
  {https://ui.adsabs.harvard.edu/abs/2019MNRAS.486.2184M} {486, 2184}

\bibitem[\protect\citeauthoryear{{Mariano} \& {Perivolaropoulos}}{{Mariano} \&
  {Perivolaropoulos}}{2012}]{Mariano12}
{Mariano} A.,  {Perivolaropoulos} L.,  2012, \mn@doi [\prd]
  {10.1103/PhysRevD.86.083517}, \href
  {https://ui.adsabs.harvard.edu/abs/2012PhRvD..86h3517M} {86, 083517}

\bibitem[\protect\citeauthoryear{{Marra}, {Amendola}, {Sawicki}  \&
  {Valkenburg}}{{Marra} et~al.}{2013}]{Marra13}
{Marra} V.,  {Amendola} L.,  {Sawicki} I.,   {Valkenburg} W.,  2013, \mn@doi
  [\prl] {10.1103/PhysRevLett.110.241305}, \href
  {http://adsabs.harvard.edu/abs/2013PhRvL.110x1305M} {110, 241305}

\bibitem[\protect\citeauthoryear{{Martinelli} \& {Tutusaus}}{{Martinelli} \&
  {Tutusaus}}{2019}]{Martinelli19}
{Martinelli} M.,  {Tutusaus} I.,  2019, \mn@doi [Symmetry]
  {10.3390/sym11080986}, 11

\bibitem[\protect\citeauthoryear{{Moffat}}{{Moffat}}{2016}]{Moffat16}
{Moffat} J.~W.,  2016, arXiv:1608.00534, \href
  {http://adsabs.harvard.edu/abs/2016arXiv160800534M} {}

\bibitem[\protect\citeauthoryear{{Mukherjee}, {Paul}  \& {Jassal}}{{Mukherjee}
  et~al.}{2019}]{Mukherjee19}
{Mukherjee} A.,  {Paul} N.,   {Jassal} H.~K.,  2019, \mn@doi [J. Cosmol.
  Astropart. Phys.] {10.1088/1475-7516/2019/01/005}, \href
  {https://ui.adsabs.harvard.edu/abs/2019JCAP...01..005M} {2019, 005}

\bibitem[\protect\citeauthoryear{{Nadathur} \& {Sarkar}}{{Nadathur} \&
  {Sarkar}}{2011}]{Nadathur11}
{Nadathur} S.,  {Sarkar} S.,  2011, \mn@doi [\prd]
  {10.1103/PhysRevD.83.063506}, \href
  {http://adsabs.harvard.edu/abs/2011PhRvD..83f3506N} {83, 063506}

\bibitem[\protect\citeauthoryear{{Odderskov}, {Koksbang}  \&
  {Hannestad}}{{Odderskov} et~al.}{2016}]{Odderskov16}
{Odderskov} I.,  {Koksbang} S.~M.,   {Hannestad} S.,  2016, \mn@doi [\jcap]
  {10.1088/1475-7516/2016/02/001}, \href
  {https://ui.adsabs.harvard.edu/abs/2016JCAP...02..001O} {2016, 001}

\bibitem[\protect\citeauthoryear{{Planck Collaboration} et~al.,}{{Planck
  Collaboration} et~al.}{2016}]{Planck16SZ}
{Planck Collaboration} et~al., 2016, \mn@doi [\aap]
  {10.1051/0004-6361/201525833}, \href
  {https://ui.adsabs.harvard.edu/abs/2016A&A...594A..24P} {594, A24}

\bibitem[\protect\citeauthoryear{{Planck Collaboration} et~al.,}{{Planck
  Collaboration} et~al.}{2018a}]{Planck18L}
{Planck Collaboration} et~al., 2018a, arXiv:1807.06205, \href
  {https://ui.adsabs.harvard.edu/abs/2018arXiv180706205P} {}

\bibitem[\protect\citeauthoryear{{Planck Collaboration} et~al.,}{{Planck
  Collaboration} et~al.}{2018b}]{Planck18}
{Planck Collaboration} et~al., 2018b, arXiv:1807.06209, \href
  {https://ui.adsabs.harvard.edu/abs/2018arXiv180706209P} {}

\bibitem[\protect\citeauthoryear{{Ramanah}, {Lavaux}, {Jasche}  \& {Wand
  elt}}{{Ramanah} et~al.}{2019}]{Ramanah19}
{Ramanah} D.~K.,  {Lavaux} G.,  {Jasche} J.,   {Wand elt} B.~D.,  2019, \mn@doi
  [\aap] {10.1051/0004-6361/201834117}, \href
  {https://ui.adsabs.harvard.edu/abs/2019A&A...621A..69R} {621, A69}

\bibitem[\protect\citeauthoryear{{Rameez}}{{Rameez}}{2019}]{Rameez19}
{Rameez} M.,  2019, arXiv:1905.00221, \href
  {https://ui.adsabs.harvard.edu/abs/2019arXiv190500221R} {}

\bibitem[\protect\citeauthoryear{{Riess} et~al.,}{{Riess}
  et~al.}{2005}]{Riess05}
{Riess} A.~G.,  et~al., 2005, \mn@doi [\apj] {10.1086/430497}, \href
  {http://adsabs.harvard.edu/abs/2005ApJ...627..579R} {627, 579}

\bibitem[\protect\citeauthoryear{{Riess} et~al.,}{{Riess}
  et~al.}{2007}]{Riess07}
{Riess} A.~G.,  et~al., 2007, \mn@doi [\apj] {10.1086/510378}, \href
  {http://adsabs.harvard.edu/abs/2007ApJ...659...98R} {659, 98}

\bibitem[\protect\citeauthoryear{{Riess} et~al.,}{{Riess}
  et~al.}{2009}]{Riess09}
{Riess} A.~G.,  et~al., 2009, \mn@doi [\apj] {10.1088/0004-637X/699/1/539},
  \href {http://adsabs.harvard.edu/abs/2009ApJ...699..539R} {699, 539}

\bibitem[\protect\citeauthoryear{{Riess} et~al.,}{{Riess}
  et~al.}{2011}]{Riess11}
{Riess} A.~G.,  et~al., 2011, \mn@doi [\apj] {10.1088/0004-637X/730/2/119},
  \href {http://adsabs.harvard.edu/abs/2011ApJ...730..119R} {730, 119}

\bibitem[\protect\citeauthoryear{{Riess} et~al.,}{{Riess}
  et~al.}{2016}]{Riess16}
{Riess} A.~G.,  et~al., 2016, \mn@doi [\apj] {10.3847/0004-637X/826/1/56},
  \href {http://adsabs.harvard.edu/abs/2016ApJ...826...56R} {826, 56}

\bibitem[\protect\citeauthoryear{{Riess}, {Casertano}, {Kenworthy}, {Scolnic}
  \& {Macri}}{{Riess} et~al.}{2018a}]{Riess18S}
{Riess} A.~G.,  {Casertano} S.,  {Kenworthy} D.,  {Scolnic} D.,   {Macri} L.,
  2018a, arXiv e-prints, \href
  {https://ui.adsabs.harvard.edu/abs/2018arXiv181003526R} {p. arXiv:1810.03526}

\bibitem[\protect\citeauthoryear{{Riess} et~al.,}{{Riess}
  et~al.}{2018b}]{Riess18}
{Riess} A.~G.,  et~al., 2018b, \mn@doi [\apj] {10.3847/1538-4357/aaadb7}, \href
  {https://ui.adsabs.harvard.edu/abs/2018ApJ...855..136R} {855, 136}

\bibitem[\protect\citeauthoryear{{Riess}, {Casertano}, {Yuan}, {Macri}  \&
  {Scolnic}}{{Riess} et~al.}{2019}]{Riess19}
{Riess} A.~G.,  {Casertano} S.,  {Yuan} W.,  {Macri} L.~M.,   {Scolnic} D.,
  2019, \mn@doi [\apj] {10.3847/1538-4357/ab1422}, \href
  {https://ui.adsabs.harvard.edu/abs/2019ApJ...876...85R} {876, 85}

\bibitem[\protect\citeauthoryear{{Rigopoulos} \& {Valkenburg}}{{Rigopoulos} \&
  {Valkenburg}}{2012}]{Rigopoulos12}
{Rigopoulos} G.,  {Valkenburg} W.,  2012, \mn@doi [\prd]
  {10.1103/PhysRevD.86.043523}, \href
  {https://ui.adsabs.harvard.edu/abs/2012PhRvD..86d3523R} {86, 043523}

\bibitem[\protect\citeauthoryear{{Schwarz}, {Copi}, {Huterer}  \&
  {Starkman}}{{Schwarz} et~al.}{2016}]{Schwarz16}
{Schwarz} D.~J.,  {Copi} C.~J.,  {Huterer} D.,   {Starkman} G.~D.,  2016,
  \mn@doi [Classical and Quantum Gravity] {10.1088/0264-9381/33/18/184001},
  \href {https://ui.adsabs.harvard.edu/abs/2016CQGra..33r4001S} {33, 184001}

\bibitem[\protect\citeauthoryear{{Scolnic} et~al.,}{{Scolnic}
  et~al.}{2018}]{Scolnic17}
{Scolnic} D.~M.,  et~al., 2018, \mn@doi [\apj] {10.3847/1538-4357/aab9bb},
  \href {https://ui.adsabs.harvard.edu/abs/2018ApJ...859..101S} {859, 101}

\bibitem[\protect\citeauthoryear{Seikel, Yahya, Maartens  \& Clarkson}{Seikel
  et~al.}{2012}]{Seikel12a}
Seikel M.,  Yahya S.,  Maartens R.,   Clarkson C.,  2012, \mn@doi [\prd]
  {10.1103/PhysRevD.86.083001}, 86, 083001

\bibitem[\protect\citeauthoryear{{Shanks}, {Hogarth}  \& {Metcalfe}}{{Shanks}
  et~al.}{2018}]{Shanks18}
{Shanks} T.,  {Hogarth} L.,   {Metcalfe} N.,  2018, arXiv e-prints, \href
  {https://ui.adsabs.harvard.edu/abs/2018arXiv181007628S} {p. arXiv:1810.07628}

\bibitem[\protect\citeauthoryear{{Shanks}, {Hogarth}  \& {Metcalfe}}{{Shanks}
  et~al.}{2019}]{Shanks19}
{Shanks} T.,  {Hogarth} L.~M.,   {Metcalfe} N.,  2019, \mn@doi [\mnras]
  {10.1093/mnrasl/sly239}, \href
  {https://ui.adsabs.harvard.edu/abs/2019MNRAS.484L..64S} {484, L64}

\bibitem[\protect\citeauthoryear{{Skillman}, {Warren}, {Turk}, {Wechsler},
  {Holz}  \& {Sutter}}{{Skillman} et~al.}{2014}]{Skillman14}
{Skillman} S.~W.,  {Warren} M.~S.,  {Turk} M.~J.,  {Wechsler} R.~H.,  {Holz}
  D.~E.,   {Sutter} P.~M.,  2014, arXiv e-prints, \href
  {https://ui.adsabs.harvard.edu/abs/2014arXiv1407.2600S} {p. arXiv:1407.2600}

\bibitem[\protect\citeauthoryear{{Soltis}, {Farahi}, {Huterer}  \&
  {Liberato}}{{Soltis} et~al.}{2019}]{Soltis19}
{Soltis} J.,  {Farahi} A.,  {Huterer} D.,   {Liberato} C.~M.,  2019, \mn@doi
  [\prl] {10.1103/PhysRevLett.122.091301}, \href
  {https://ui.adsabs.harvard.edu/abs/2019PhRvL.122i1301S} {122, 091301}

\bibitem[\protect\citeauthoryear{Stahl}{Stahl}{2016}]{Stahl16}
Stahl C.,  2016, \mn@doi [International Journal of Modern Physics D]
  {10.1142/S0218271816500668}, 25, 1650066

\bibitem[\protect\citeauthoryear{{Sun} \& {Wang}}{{Sun} \&
  {Wang}}{2018}]{Sun18}
{Sun} Z.~Q.,  {Wang} F.~Y.,  2018, \mn@doi [\mnras] {10.1093/mnras/sty1391},
  \href {https://ui.adsabs.harvard.edu/abs/2018MNRAS.478.5153S} {478, 5153}

\bibitem[\protect\citeauthoryear{{Szapudi} et~al.,}{{Szapudi}
  et~al.}{2015}]{Szapudi15}
{Szapudi} I.,  et~al., 2015, \mn@doi [\mnras] {10.1093/mnras/stv488}, \href
  {https://ui.adsabs.harvard.edu/abs/2015MNRAS.450..288S} {450, 288}

\bibitem[\protect\citeauthoryear{{Taubenberger} et~al.,}{{Taubenberger}
  et~al.}{2019}]{Taubenberger19}
{Taubenberger} S.,  et~al., 2019, \mn@doi [\aap] {10.1051/0004-6361/201935980},
  \href {https://ui.adsabs.harvard.edu/abs/2019A&A...628L...7T} {628, L7}

\bibitem[\protect\citeauthoryear{{Tokutake} \& {Yoo}}{{Tokutake} \&
  {Yoo}}{2016}]{Tokutake16}
{Tokutake} M.,  {Yoo} C.-M.,  2016, \mn@doi [J. Cosmol. Astropart. Phys.]
  {10.1088/1475-7516/2016/10/009}, \href
  {https://ui.adsabs.harvard.edu/\#abs/2016JCAP...10..009T} {2016, 009}

\bibitem[\protect\citeauthoryear{{Tokutake}, {Ichiki}  \& {Yoo}}{{Tokutake}
  et~al.}{2018}]{Tokutake17}
{Tokutake} M.,  {Ichiki} K.,   {Yoo} C.-M.,  2018, \mn@doi [\jcap]
  {10.1088/1475-7516/2018/03/033}, \href
  {https://ui.adsabs.harvard.edu/abs/2018JCAP...03..033T} {2018, 033}

\bibitem[\protect\citeauthoryear{{Tolman}}{{Tolman}}{1934}]{Tolman34}
{Tolman} R.~C.,  1934, \mn@doi [Proc. Natl. Acad. Sci. U.S.A.]
  {10.1073/pnas.20.3.169}, \href
  {http://adsabs.harvard.edu/abs/1934PNAS...20..169T} {20, 169}

\bibitem[\protect\citeauthoryear{{Tutusaus}, {Lamine}  \&
  {Blanchard}}{{Tutusaus} et~al.}{2019}]{Tutusaus19}
{Tutusaus} I.,  {Lamine} B.,   {Blanchard} A.,  2019, \mn@doi [\aap]
  {10.1051/0004-6361/201833032}, \href
  {https://ui.adsabs.harvard.edu/abs/2019A&A...625A..15T} {625, A15}

\bibitem[\protect\citeauthoryear{{Valkenburg}}{{Valkenburg}}{2012}]{Valkenburg12}
{Valkenburg} W.,  2012, \mn@doi [General Relativity and Gravitation]
  {10.1007/s10714-012-1405-9}, \href
  {https://ui.adsabs.harvard.edu/\#abs/2012GReGr..44.2449V} {44, 2449}

\bibitem[\protect\citeauthoryear{{Valkenburg}, {Marra}  \&
  {Clarkson}}{{Valkenburg} et~al.}{2014}]{Valkenburg14}
{Valkenburg} W.,  {Marra} V.,   {Clarkson} C.,  2014, \mn@doi [\mnras]
  {10.1093/mnrasl/slt140}, \href
  {https://ui.adsabs.harvard.edu/abs/2014MNRAS.438L...6V} {438, L6}

\bibitem[\protect\citeauthoryear{{Vargas}, {Falciano}  \& {Reis}}{{Vargas}
  et~al.}{2017}]{Vargas17}
{Vargas} C.~Z.,  {Falciano} F.~T.,   {Reis} R.~R.~R.,  2017, \mn@doi [Classical
  and Quantum Gravity] {10.1088/1361-6382/34/2/025002}, \href
  {http://adsabs.harvard.edu/abs/2017CQGra..34b5002V} {34, 025002}

\bibitem[\protect\citeauthoryear{{Wang} \& {Wang}}{{Wang} \&
  {Wang}}{2014}]{Wang14}
{Wang} J.~S.,  {Wang} F.~Y.,  2014, \mn@doi [\mnras] {10.1093/mnras/stu1279},
  \href {https://ui.adsabs.harvard.edu/abs/2014MNRAS.443.1680W} {443, 1680}

\bibitem[\protect\citeauthoryear{{Whitbourn} \& {Shanks}}{{Whitbourn} \&
  {Shanks}}{2014}]{Whitbourn14}
{Whitbourn} J.~R.,  {Shanks} T.,  2014, \mn@doi [\mnras]
  {10.1093/mnras/stt2024}, \href
  {https://ui.adsabs.harvard.edu/abs/2014MNRAS.437.2146W} {437, 2146}

\bibitem[\protect\citeauthoryear{{Whitbourn} \& {Shanks}}{{Whitbourn} \&
  {Shanks}}{2016}]{Whitbourn16}
{Whitbourn} J.~R.,  {Shanks} T.,  2016, \mn@doi [\mnras]
  {10.1093/mnras/stw555}, \href
  {https://ui.adsabs.harvard.edu/abs/2016MNRAS.459..496W} {459, 496}

\bibitem[\protect\citeauthoryear{{Wojtak} \& {Prada}}{{Wojtak} \&
  {Prada}}{2017}]{Wojtak17}
{Wojtak} R.,  {Prada} F.,  2017, \mn@doi [\mnras] {10.1093/mnras/stx1550},
  \href {http://adsabs.harvard.edu/abs/2017MNRAS.470.4493W} {470, 4493}

\bibitem[\protect\citeauthoryear{{Wojtak}, {Knebe}, {Watson}, {Iliev},
  {He{\ss}}, {Rapetti}, {Yepes}  \& {Gottl{\"o}ber}}{{Wojtak}
  et~al.}{2014}]{Wojtak14}
{Wojtak} R.,  {Knebe} A.,  {Watson} W.~A.,  {Iliev} I.~T.,  {He{\ss}} S.,
  {Rapetti} D.,  {Yepes} G.,   {Gottl{\"o}ber} S.,  2014, \mn@doi [\mnras]
  {10.1093/mnras/stt2321}, \href
  {https://ui.adsabs.harvard.edu/abs/2014MNRAS.438.1805W} {438, 1805}

\bibitem[\protect\citeauthoryear{{Wong} et~al.,}{{Wong} et~al.}{2019}]{Wong19}
{Wong} K.~C.,  et~al., 2019, arXiv:1907.04869, \href
  {https://ui.adsabs.harvard.edu/abs/2019arXiv190704869W} {}

\bibitem[\protect\citeauthoryear{{Wu} \& {Huterer}}{{Wu} \&
  {Huterer}}{2017}]{Wu17}
{Wu} H.-Y.,  {Huterer} D.,  2017, \mn@doi [\mnras] {10.1093/mnras/stx1967},
  \href {https://ui.adsabs.harvard.edu/abs/2017MNRAS.471.4946W} {471, 4946}

\bibitem[\protect\citeauthoryear{{Zhang}, {Zhang}, {Wang}  \& {Ma}}{{Zhang}
  et~al.}{2015}]{Zhang15}
{Zhang} Z.-S.,  {Zhang} T.-J.,  {Wang} H.,   {Ma} C.,  2015, \mn@doi [\prd]
  {10.1103/PhysRevD.91.063506}, \href
  {http://adsabs.harvard.edu/abs/2015PhRvD..91f3506Z} {91, 063506}

\bibitem[\protect\citeauthoryear{{Zhang} et~al.,}{{Zhang}
  et~al.}{2019}]{Zhang19}
{Zhang} Z.,  et~al., 2019, \mn@doi [\apj] {10.3847/1538-4357/ab1ea4}, \href
  {https://ui.adsabs.harvard.edu/abs/2019ApJ...878..137Z} {878, 137}

\bibitem[\protect\citeauthoryear{{Zhao} et~al.,}{{Zhao} et~al.}{2019a}]{Zhao19}
{Zhao} G.-B.,  et~al., 2019a, \mn@doi [\mnras] {10.1093/mnras/sty2845}, \href
  {https://ui.adsabs.harvard.edu/abs/2019MNRAS.482.3497Z} {482, 3497}

\bibitem[\protect\citeauthoryear{{Zhao}, {Zhou}  \& {Chang}}{{Zhao}
  et~al.}{2019b}]{ZhaoD19}
{Zhao} D.,  {Zhou} Y.,   {Chang} Z.,  2019b, \mn@doi [\mnras]
  {10.1093/mnras/stz1259}, \href
  {https://ui.adsabs.harvard.edu/abs/2019MNRAS.486.5679Z} {486, 5679}

\bibitem[\protect\citeauthoryear{{Zibin}}{{Zibin}}{2008}]{Zibin08}
{Zibin} J.~P.,  2008, \mn@doi [\prd] {10.1103/PhysRevD.78.043504}, \href
  {http://adsabs.harvard.edu/abs/2008PhRvD..78d3504Z} {78, 043504}

\bibitem[\protect\citeauthoryear{{Zibin}}{{Zibin}}{2011}]{Zibin11}
{Zibin} J.~P.,  2011, \mn@doi [\prd] {10.1103/PhysRevD.84.123508}, \href
  {http://adsabs.harvard.edu/abs/2011PhRvD..84l3508Z} {84, 123508}

\bibitem[\protect\citeauthoryear{{du Mas des Bourboux} et~al.,}{{du Mas des
  Bourboux} et~al.}{2017}]{MasdesBourboux17}
{du Mas des Bourboux} H.,  et~al., 2017, \mn@doi [\aap]
  {10.1051/0004-6361/201731731}, \href
  {http://adsabs.harvard.edu/abs/2017A%26A...608A.130D} {608, A130}

\makeatother
\end{thebibliography}
\bibliographystyle{mnras}
\appendix
\section{Top-hat density profile}
\label{sec:app}
\begin{figure*}
\includegraphics[width=\textwidth]{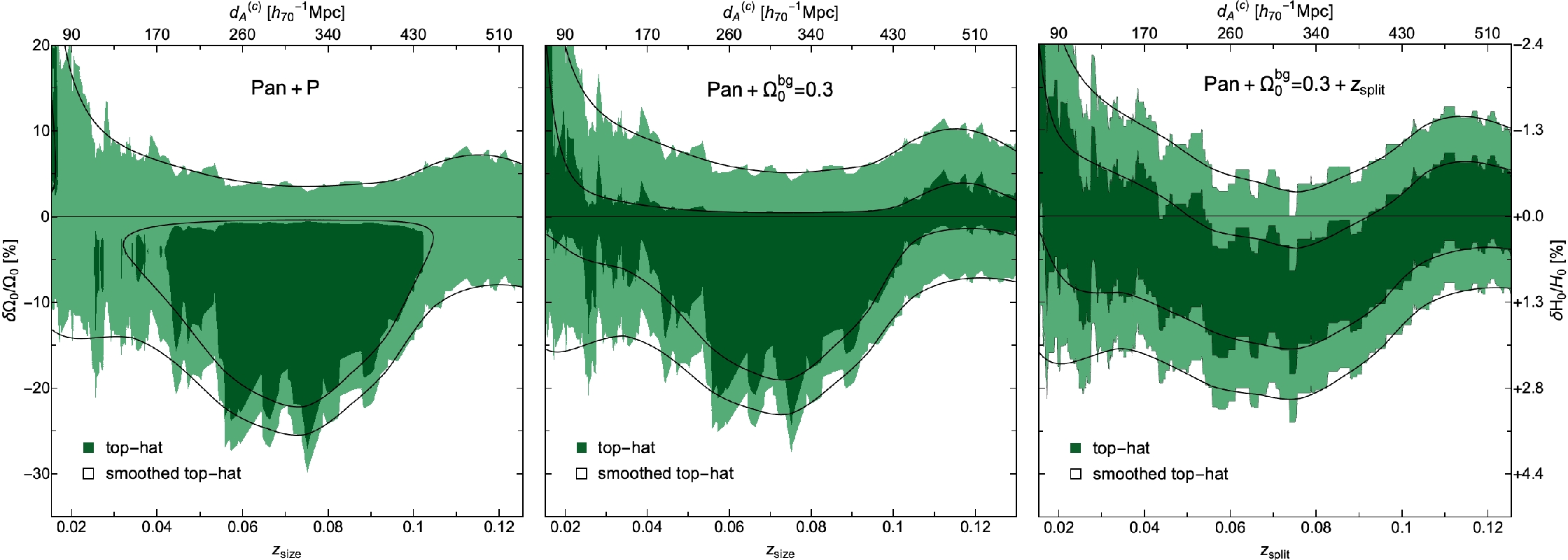}
\caption{$1\sigma$ and $2\sigma$ confidence regions for $\Lambda$LTB model with TH density profile, fitted to Pantheon dataset are shown as coloured contours, while the solid lines represent the confidence regions obtained from the smoothed $\chi^2$ function. In the first panel we used Planck prior for $\Omega_m^{\rm bg}$, whereas the second and third panels are obtained with fixed $\Omega_m^{\rm bg}=0.3$. Moreover, in the third case $z_{\rm size}$ parameter was a priori fixed to a range of $z_{\rm split}$ values, followed by $\chi^2$ sampling over $\delta\Omega_0/\Omega_0$ contrast values.}
\label{fig:THSPan}
\end{figure*}
Our results for $\Lambda$LTB model can be reproduced to a good extent using a simpler TH density profile constructed as a two-step homogeneous $k\Lambda$CDM model.
The following approximate formulae are derived from $\Lambda$LTB model as third order series expansion in terms of present matter density contrast $x=\delta\rho_0/\rho_0$.
Given that $\Lambda$ is constant, its dimensionless density parameter is not (see \cref{sec:th}).
\ba
{\delta H_0\over H_0}=&x\left(-0.17093-0.32158 (\Omega_m^{\rm bg}-0.3)+0.24932(\Omega_m^{\rm bg}-0.3)^2\right)\nonumber\\
+&x^2\left(0.03141+0.063 (\Omega_m^{\rm bg}-0.3)\right)\nonumber\\
+&x^3(-0.02237)\label{eq:SEH}
\\
{\delta\Omega_0\over\Omega_0}=&x\left(1.34186+0.64317 (\Omega_m^{\rm bg}-0.3)-0.49863(\Omega_m^{\rm bg}-0.3)^2\right)\nonumber\\
+&x^2\left(0.36669+0.84699 (\Omega_m^{\rm bg}-0.3)\right)\nonumber\\
+&x^3(+0.0563)\label{eq:SEM}
\\
{\delta\Omega_\Lambda\over\Omega_\Lambda}=&x\left(0.34186+0.64317(\Omega_m^{\rm bg}-0.3)-0.49863(\Omega_m^{\rm bg}-0.3)^2\right)\nonumber\\
+&x^2\left(0.02483+0.20382 (\Omega_m^{\rm bg}-0.3)\right)\nonumber\\
+&x^3(+0.03211)\label{eq:SEL}
\end{align}
In the range $0.2<\Omega_m^{\rm bg}<0.4$ and $-30\%<\delta\rho_0<30\%$ these eqs. are correct with less than $1\%$ error on the value.
The constructed TH model can be characterised with three cosmological parameters: $\Omega_m^{\rm bg}$, $\delta\Omega_0/\Omega_0$ and $z_{\rm size}$, as well as the intercept SN Ia magnitude parameter $\cal M$.
As usual, using a prior on $\Omega_m^{\rm bg}$ ensures tighter constraints on void parameters.
Using \cref{eq:SEH,eq:SEM,eq:SEL} together with two-step $k\Lambda$CDM formalism, it is easy to evaluate all the distances of SN Ia inside and outside the void.

Sampling Pantheon $\chi^2$ for different model parameters' values, we immediately notice it can be very irregular along the redshift axis, since the top-hat profile does not have a smooth transition from local to background geometry (see \Cref{fig:THSPan}).
We find that a TH void with a contrast $\delta\Omega_0/\Omega_0$\,$\approx$\,$-25\%$ has comparable $\chi^2$ to the best-fit of $\Lambda$CDM model for specifically chosen values of $z_{\rm size}\approx0.075$, although this fitting method has no way of accounting for the systematic error arousing with fixing the physical size of the void.
To overcome this, one can perform smoothing of the sampled TH likelihood over redshift bins of size $\Delta z=0.01$.
Although short in size, these bins have $\sim$\,$20$ low-redshift SN each, regularising the average likelihood dependence on $z_{\rm size}$.
The $1\sigma$ and $2\sigma$ confidence regions resulting from smoothed likelihoods are shown as solid lines in \Cref{fig:THSPan}.
We note that contours obtained in this way for Pan+P data are substantially similar to what we got using the GBH profile (c.f. \Cref{fig:GBHPanKBC}).
After properly marginalising out other parameters in the smoothed likelihood, we get a Pan+P constraint on contrast as $(\delta\Omega_0/\Omega_0)^{c.i.}=-9.0_{-6.8}^{+7.6}\%$.
The corresponding result in \Cref{tab:1} has somewhat wider error bars since the smooth GBH function allows for more freedom in density profile than the TH form.

The second and third panels in \Cref{fig:THSPan} show the importance of fitting the model parameters in this scheme as opposed to fixing them.
Specifically, the last panel is obtained by
sampling Pantheon $\chi^2$ for contrast $\delta\Omega_0/\Omega_0$ and $\cal M$ parameters, while fixing the void size to $z_{\rm split}$ and $\Omega_0^{\rm bg}=0.3$.
Hence, this method is unable to provide a constraint on void size, just as the constraint on density contrast depends on the prior choice of $z_{\rm split}$.
The contours obtained in third panel of \Cref{fig:THSPan} are sequences of confidence intervals for $\delta\Omega_0/\Omega_0$ parameter,
recognised by a stripe shape in $z_{\rm split}-\delta\Omega_0/\Omega_0$ plane.

\label{lastpage}
\end{document}